\documentclass[a4paper,11pt]{article}

\usepackage{amssymb}
\usepackage{amsmath}
\usepackage{amscd}
\usepackage{latexsym}
\usepackage{epsfig}
\usepackage{color}
\usepackage{fancyhdr}
\usepackage[bf,footnotesize]{caption}
\usepackage{graphicx}
\usepackage{booktabs}
\usepackage{a4wide}

\usepackage[dvips,colorlinks=true,a4paper=true,backref=false, linktocpage=true,
citecolor=blue,urlcolor=blue,linkcolor=blue,pdfpagemode=UseOutlines]{hyperref}

\hypersetup{%
  bookmarksnumbered=true,
  pdftitle = {},
  pdfsubject = {},
  pdfauthor = {U.~Wenger},
  pdfkeywords = {}
}

\graphicspath{{Figures-small/}}

\def\det{\mathop{\rm det}}              
\def\Dslash{\hbox{D}\kern-0.6em\raise0.15ex\hbox{/}} 

\def\Z{{\mathbb Z}}                     

\def\onehalf{\frac{1}{2}}               

\def\bi{\begin{itemize}}
\def\ei{\end{itemize}}
\def\be{\begin{equation}}
\def\ee{\end{equation}}

\def\Umat{{\cal U}}

\date{\empty}

\begin{document}
\title{QCD at non-zero density and canonical partition functions with Wilson fermions}

\author{Andrei Alexandru$^{a}$  and Urs Wenger$^b$\vspace{0.5cm}
\\
$^a\,$Physics Department, The George Washington University \\
Washington, DC 20052, USA \\
\\
$^b\,$Albert Einstein Center for Fundamental Physics\\
Institute for Theoretical Physics, Sidlerstrasse 5,
CH--3012 Bern, Switzerland\vspace{0.5cm}
\\
}

\maketitle

\begin{abstract}
\noindent
We present a reduction method for Wilson Dirac fermions with non-zero
chemical potential which generates a dimensionally reduced fermion
matrix. The size of the reduced fermion matrix is independent of the
temporal lattice extent and the dependence on the chemical potential
is factored out. As a consequence the reduced matrix allows a simple
evaluation of the Wilson fermion determinant for any value of the
chemical potential and hence the exact projection to the canonical
partition functions.
\end{abstract}

\section{Introduction}
Non-perturbative lattice calculations of Quantum Chromodynamics (QCD)
at zero density have seen remarkable progress in recent years.
However, simulations at non-zero quark or baryon density remain a
challenge due to the occurrence of a complex phase in the fermion
determinant at non-zero chemical potential. The fluctuation of this
phase is the source of the notorious fermionic sign problem and
obstructs the straightforward simulation of the theory
using Monte-Carlo importance sampling. This problem limits the
reliability of present day lattice QCD calculations at finite baryon density and
makes it difficult to explore the QCD phase diagram in parameter
regimes which are particularly interesting, e.g.~for the
identification of different phases of matter, the determination of
phase transition lines and the location of possible critical
endpoints. However, credible non-perturbative results at non-zero
quark density would provide important phenomenological information,
e.g.~for understanding the structure of neutron stars or the dynamics
of relativistic heavy-ion collisions.

One approach to QCD at finite density makes use of the
canonical formulation where the net quark (or baryon) number is
held constant. This can be achieved by separating the grand-canonical
partition function $Z_{GC}$ into a sum of canonical partition functions
$Z_C(k)$ with a fixed net number $k$ of quarks and anti-quarks.  Quantities at fixed
chemical potential, i.e.~in the standard grand-canonical formulation
of QCD, can then be obtained by averaging over the canonical partition
functions. It
turns out that the projection to the canonical sectors can be done
exactly, gauge field by gauge field, however it requires the
integration of the fermion determinant over the whole range of
imaginary chemical potential $\phi=i\mu/T \in [0,2\pi]$.

In the past this approach could only be made practical in connection
with staggered fermions. For those, clever fermion matrix reduction
methods were developed
\cite{Gibbs:1986hi,Barbour:1988ax,Hasenfratz:1991ax} that allow the
evaluation of the determinant for any value of the chemical potential
once the eigenvalues of the reduced fermion matrix are known. In this
approach the reduction of the fermion matrix in size by a factor half
the temporal lattice extent is the crucial ingredient
since it reduces the complexity of the eigenvalue computation by the
corresponding factor cubed. Unfortunately, however, for Wilson
fermions so far no such reduction method was known despite various
attempts \cite{Danzer:2008xs,Gattringer:2009wi}.

In this paper we present such a reduction method for Wilson fermions,
i.e.~we derive a dimensionally reduced Wilson fermion matrix whose
size is independent of the temporal lattice extent and for which the
dependence on the chemical potential is factored out. It therefore
allows easy and exact evaluation of the determinants at any value of
the chemical potential and hence the straightforward projection to the
various canonical sectors. Applications which are facilitated by the
reduction method for Wilson fermions presented here include the
reweighting of ensembles to different values of the chemical potential
and calculations based on canonical ensembles
\cite{Kratochvila:2005mk,Alexandru:2005ix,Fodor:2007ga,deForcrand:2007uz}.

The reduction of the four dimensional Wilson Dirac operator to
the three dimensional reduced fermion matrix is very similar to the
construction of the four dimensional overlap operator from the five
dimensional domain wall fermion operator
\cite{Borici:2004pn,Edwards:2005an,Edwards:2010}.  A similar reduction method for
the Wilson fermion matrix has also been proposed in
\cite{Borici:2004bq} in the context of reweighting with stochastic
determinants. Finally, while preparing the paper we were informed by
Nakamura and Nagata \cite{nakamura:2010} about their development of
similar reduction techniques for the Wilson fermion matrix.

The paper is organized as follows. In section
\ref{sec:canonicalFormulation} we briefly review the separation of the
grand-canonical partition function of QCD into a sum of canonical
partition functions with fixed quark or baryon number. In section
\ref{sec:WilsonReduction} we present the reduction method for Wilson
fermions which renders the computational complexity of the determinant
independent of the temporal lattice extent and factorizes the
dependence on the chemical potential. In section
\ref{sec:spectral_properties} and \ref{sec:ReductionResults} we
discuss spectral properties of the reduced matrix and some properties
of the projected determinants, respectively. While the results from
these sections so far do not have a direct physical application, we
would like to emphasise their potential importance for the development of new
canonical simulation algorithms, or for the optimization of
reweighting strategies.  Finally, in section
\ref{sec:canonicalResults} we present some results from a reweighting
of canonical ensembles, merely as a demonstration of the potential of
the reduced fermion matrix approach.

\section{Canonical formulation of QCD at fixed baryon number}
\label{sec:canonicalFormulation}
The grand-canonical partition function at temperature $T$ and chemical
potential $\mu_q$ for a single quark flavor can be defined as
\begin{equation}\label{eq:def_ZGC}
Z_{GC}(\mu_q) = \int {\cal D}U \, e^{-S_{g}(U)}{\det} M(U,\mu_q),
\end{equation}
where $M(U,\mu_q)$ denotes the Dirac operator, $U$ collects the gauge
field degrees of freedom from the color gauge group $\text{SU}(N_c)$ and
$S_{g}(U)$ is the gauge field action.  This is the commonly used
partition function for simulating QCD thermodynamics on the lattice
\cite{Fodor:2001au,Allton:2002zi,deForcrand:2002ci,D'Elia:2002gd},
which in general, however, suffers from a strong fermionic sign
problem. The same thermodynamic physics can also be extracted using
the canonical partition function
\cite{Kratochvila:2005mk,Alexandru:2005ix,Kratochvila:2004wz,Alexandru:2004dx,Ejiri:2008xt},
although one should keep in mind that the physics of the two systems
strictly coincide only in the thermodynamic limit, i.e.~in the limit
of infinite spatial volume.  The partition function in the canonical
approach, for a system with a net number of $k$ quarks, can be written
as
\begin{equation}\label{eq:def_ZC}
Z_{C}(k) = \int {\cal D}U \, e^{-S_{g}(U)}{\det}_{k} M(U),
\end{equation}
where the fermionic contribution is now included in the projected
determinant
\begin{equation}\label{eq:def_detk}
{\det}_{k} M(U) \equiv \frac{1}{2\pi}\int_{0}^{2\pi} d\phi \,
e^{-ik\phi}  \det M(U, \mu_q = i \phi T)\,
\end{equation}
and one has made use of the fact that $\det M(U, \mu_q = i \phi T)$
enjoys a $\frac{2\pi}{N_c}$-periodicity in $\phi$
\cite{Roberge:1986mm,Kratochvila:2006jx}.  The periodicity stems from
the fact that a shift in the imaginary chemical potential $\phi
\rightarrow \phi+\frac{2\pi}{N_c}$ can be exactly compensated by a
corresponding $\Z(N_c)$-transformation of the underlying gauge field.
From this periodicity it also follows that $Z_{C}(k) = 0$ for $k \neq
0\! \mod N_c$, i.e.~it vanishes for non-integer baryon numbers $n_B
\in \hspace{-2.1ex} \slash \hspace{1.5ex}\Z$, while $Z_{C}(k) =
Z_{C}^*(-k)$ follows from the evenness $Z_{GC}(\mu_q) =
Z_{GC}(-\mu_q)$ due to time-reversal symmetry.

Finally, one can relate the canonical partition functions back to the
grand-canonical ones using the fugacity expansion
\be \label{eq:fugacity_expansion}
Z_{GC}(\mu_q) = \sum_{k=-\infty}^{+\infty} e^{k \mu_q/T} Z_C(k) \, ,
\ee
where the sum can in principle be restricted to $k = 0\! \mod N_c$,
i.e.~integer baryon numbers, following the discussion
above. Furthermore, the equation also motivates the determination of
the baryon chemical potential in the canonical approach by a
definition based on the free energy. Essentially, the baryon chemical
potential is the response of the system when introducing one more
baryon to the system, i.e.
\begin{equation}\label{eq:muB_def}
\mu_B(n_B) \equiv F(N_c\cdot(n_B+1)) - F(N_c \cdot n_B) \,,
\end{equation}
where $F(k) = -T \log Z_C(k)$ is the Helmholtz free energy of the
canonical partition function. In a finite volume $V$, this definition is
ambiguous due to the discreteness of the baryon number, however, in
the thermodynamic limit it yields $\mu_{B}(\rho_B) = df/d\rho_B$,
where $\rho_B=n_B/V$ and $f=F/V$ are the baryon and free energy
densities.  Note that the baryon chemical potential $\mu_{B}$ is
different from the quark chemical potential $\mu_{q}$ used above. The
quark chemical potential cannot be defined as the increase of the free
energy when introducing a quark in the system since the free energy is
infinite for systems that have fractional baryon numbers.  If we need
to compare the chemical potential used in the grand-canonical approach
to the chemical potential measured in the canonical approach, we use
$\mu_{q} \simeq \mu_{B}/N_{c}$.

\section{Reduction technique for the Wilson fermion matrix}
\label{sec:WilsonReduction}
In the following we consider QCD with Wilson fermions on a periodic
lattice with temporal and spatial extent $L_t$ and $L_s$,
respectively. The quark chemical potential is $\mu \equiv a \mu_q$
where $a$ is the lattice spacing.\footnote{From now on we set $a=$1.}
The massive Wilson-Dirac operator can be written as
\begin{equation}\label{eq:DiracMatrix}
     M = \onehalf\Gamma_\nu(\nabla_\nu+\nabla_\nu^*)
   - \onehalf\nabla_\nu^*\nabla_\nu + m,
\end{equation}
where $\nabla_\nu,\,\nabla_\nu^*$ denote the covariant forward and
backward lattice derivative, $\Gamma_\nu$ are the Euclidean Dirac
matrices and $m$ is the bare quark mass. The
chemical potential $\mu$ couples to the fermion number operator and is
introduced on the lattice \cite{Hasenfratz:1983ba} by furnishing the
forward and backward temporal hopping terms by factors of
$e^{\pm\mu}$, respectively. More explicitly we have
\begin{multline}\label{eq:M_explicit}
M_{x,y} = (m + 4 ) \cdot \delta_{x,y} - \sum_{k=1}^3 \left\{ P(+k) \,
  U_k(x) \,
\delta_{y,x+\hat k} + P(-k) \, U_k^\dagger(y) \, \delta_{y,x-\hat k}
\right\}  \\
-   \left\{ e^{+\mu} P(+4) \, U_4(x) \,
\delta_{y,x+\hat 4} +  e^{-\mu} P(-4) U_4^\dagger(y) \, \delta_{y,x-\hat 4} \right\},
\end{multline}
where $U_\nu(x) \in \text{SU}(N_c)$ are the gauge field links and
$P(\pm \nu) = \frac{1}{2}(1\mp\Gamma_\nu), \,
\nu=1,\ldots,4$. Further, for convenience we introduce $P_\pm \equiv
P(\pm 4)$ for the projectors in temporal direction and will use this
notation in the following. Choosing the spatial Euclidean Dirac matrices as
\be \label{eq:Gamma_matrices}
\Gamma_k = \left(
\begin{array}{cc}
0 & \sigma_k \\
\sigma_k^\dagger & 0
\end{array}
\right) 
\ee
and hence hermitian, we note that the spatial part of the Wilson Dirac
operator, i.e.~the first line of eq.(\ref{eq:M_explicit}), can be
written in the form
\be \label{eq:spatial_B}
B = \left( 
\begin{array}{cc}
B_{++} & C \\
-C^\dagger & B_{--}
\end{array}
 \right)
\ee
where $B_{++} = B_{--}$ is hermitian and trivial in Dirac space,
\be\label{eq:Bdef}
(B_{++})_{x,y} = (m + 4) \cdot \delta_{x,y} - \frac{1}{2} \sum_{k=1}^3 \left\{\delta_{y,x+\hat k} U_k(x) + \delta_{x,y+\hat k} U_k^\dagger(y)\right\},
\ee
while
\be\label{eq:Cdef}
C_{x,y} = \frac{1}{2} \sum_{k=1}^3 \sigma_k \left\{ \delta_{y,x+\hat
    k}  U_k(x) - \delta_{x,y+\hat k} U_k^\dagger(y) \right\}
\ee
is hermitian if the Heisenberg matrices $\sigma_k$ in
eq.(\ref{eq:Gamma_matrices}) are chosen to be hermitian.  The
derivations presented below focus on the un-improved version of the
Wilson Dirac operator. However, the addition of the clover term can be
easily accommodated: it only changes the spatial hopping matrix $B$ in
a way that is consistent with the structure in eq.(\ref{eq:spatial_B})
which is all that is needed for some of the more specific derivations in sections
\ref{sec:calculationofT} and \ref{sec:TU_symmetry}. In fact, the
numerical experiments presented later in the paper use the clover
improved version of the Dirac operator.

\subsection{Reduction of $\det M$}\label{sec:detM_reduction}
The Wilson fermion matrix (\ref{eq:DiracMatrix}) in temporal gauge
with (anti-)periodic boundary conditions in space (time)direction for
a single quark flavor with chemical potential $\mu$ can be represented
by
\begin{equation}\label{eq:tildeM0}
 M = \left(  
\begin{array}{ccccc}
 B_0  & P_+  &        &        &  -P_- \cdot \Umat^\dagger \cdot e^{-\mu L_t} \\
 P_-  & B_1  & P_+    &        &  \\
      & P_-  & B_2    & \ddots &   \\
      &      & \ddots & \ddots &   \\
      &      &        &        &  P_+ \\
-P_+ \cdot \Umat \cdot e^{+\mu L_t} & & & P_-    &  B_{L_t-1} \\
\end{array}
\right)
\end{equation}
where the $B_t$'s are $(4\cdot N_c \cdot L_s \times 4\cdot N_c\cdot
L_s)$-matrices and represent the (spatial) Wilson Dirac operator on
time-slice $t$ and  where we have rescaled the fermion fields such that the
dependence on the chemical potential resides on the links connecting
the last and the first time slice. The temporal gauge links also
reside only on those links and are collected in the matrix $\Umat$ so
that $P_+ \cdot \Umat$ is of the same size as $B_t$. For convenience
we abbreviate in the following $A^- \equiv - \Umat^\dagger \cdot
e^{-\mu L_t}$ and $A^+ \equiv - \Umat \cdot e^{+\mu L_t}$ and note
that the matrix can be written in the form
\begin{equation}\label{eq:tildeM}
 M = 
\left(  
\begin{array}{cccccc}
 B_0 &       &        &  A^- \\
 1   & B_1   &        &      \\
     &\ddots & \ddots &      \\
     &       &    1   &  B_{L_t-1} 
\end{array}
\right) P_- \, + \,
\left(  
\begin{array}{cccccc}
 B_0 & 1   &        & \\
     & B_1 & \ddots & \\
     &     & \ddots & 1  \\
 A^+ &     &        &  B_{L_t-1} 
\end{array}
\right) P_+
\end{equation}
using the fact that $P_+ + P_- = 1$. Essentially, this splits the
matrix into two parts describing the components of the Dirac particle
moving forward and backward in time. Next we define the
shift-projection matrix
\begin{equation}
 {\cal P} = \left(  
\begin{array}{cccccc}
P_+ & P_- &     &        &  \\
    & P_+ & P_- &        &  \\
    &     & P_+ & \ddots &   \\
    &     &     & \ddots & P_-  \\
P_- &    &     &        & P_+ 
\end{array}
\right) 
\end{equation}
which leaves the forward moving part invariant and shifts the backward
moving part by on time slice. We further note that $\det {\cal P} =
1$. Multiplying $M$ with ${\cal P}$ from the right we find
\begin{eqnarray}
 M \cdot {\cal P} &=& 
\left(  
\begin{array}{cccccc}
Q_0^- (P_- A^- + P_+)&  Q_0^+ &       &          &             \\ 
                    &  Q_1^- & Q_1^+ &          &              \\
                    &       & Q_2^-  & \ddots  &              \\
                    &       &       & \ddots   &  Q_{L_t-2}^+   \\
Q_{L_t-1}^+  (P_+ A^+ + P_-)&       &       &          &  Q_{L_t-1}^- 
\end{array}
\right)
\end{eqnarray}
where we defined
\begin{equation}
 Q_i^\pm = B_i P_\mp + P_\pm \, 
\end{equation}
and used the fact that 
\begin{eqnarray}
P_- A^- + B_0 P_+ &=& Q_0^- (P_- A^- + P_+) , \\
P_+ A^+ + B_{L_t-1} P_- &=& Q_{L_t-1}^+ (P_+ A^+ + P_-) .
\end{eqnarray}
Now we define the block diagonal matrix
\begin{equation}
 {\cal Q} = \left(  
\begin{array}{cccc}
Q_0^- &       &        &            \\
      & Q_1^- &        &            \\
      &       & \ddots &            \\
      &       &        & Q_{L_t-1}^- \\
\end{array}
\right) 
\end{equation}
and find
\begin{equation}\label{eq:transfer_matrix}
\tilde M \equiv {\cal Q}^{-1} \cdot M \cdot {\cal P} =
\left(  
\begin{array}{ccccc}
(P_- A^- + P_+)           &  T_0 &        &          &         \\ 
                         &  1   & T_1    &          &         \\
                         &      & 1      &  \ddots  &         \\
                         &      &        &  \ddots  & T_{L_t-2} \\
 T_{L_t-1} (P_+ A^+ + P_-)  &      &        &          &    1    
\end{array}
\right)
\end{equation}
where
\begin{equation}
T_i = (Q_i^-)^{-1} \cdot Q_i^+ \,.
\end{equation}
Note that the matrix $\tilde M$ is essentially a transfer matrix
describing fermions hopping forward and backward between time
slices. We discuss this further in section
\ref{sec:physical_interpretation}.

We can now easily calculate the determinant of the transfer matrix
$\tilde M$ using Schur complement techniques
\cite{Edwards:2010}. Defining
\begin{equation}
T \equiv  T_0  \cdot T_1\cdot \ldots \cdot T_{L_t-1}
\end{equation}
we find
\begin{equation}
\det \left[ {\cal Q}^{-1} \cdot M \cdot {\cal P} \right] = 
\det \left[ (P_- A^- + P_+) - (-1)^{L_t}T \cdot (P_+ A^+ + P_-)\right] 
\label{eq:reduced determinant 1}
\end{equation}
and hence 
\begin{equation}
\det \left[ M \right] = \left(\prod_{i=0}^{L_t-1} \det Q_i^- \right) \cdot 
\det \left[ {\cal Q}^{-1} \cdot M \cdot {\cal P} \right]\, .
\end{equation}
Note that the first factor $\det[{\cal Q}]$ from canceling the effect
of multiplication with $ {\cal Q}^{-1} $ is independent of $\mu$ and
cancels when we take the ratio of two determinants, e.g.~with two
different chemical potentials. Furthermore, from now on we assume
$L_t$ to be even in order to get rid of the inconvenient factor
$(-1)^{L_t}$.

In order to separate the dependence on the chemical potential $\mu$
from the gauge field dependence we first note that for arbitrary
matrices $A, B, C, D$ diagonal in Dirac space, i.e.~commuting with
$P_\pm$, we have
\begin{equation}
(P_\pm A + P_\mp B) (P_\pm C + P_\mp D) = P_\pm A C + P_\mp B D \,.
\end{equation}
So multiplying eq.(\ref{eq:reduced determinant 1}) by $\det[P_+ A + P_- B]$
with $A = e^{-\mu L_t}$ and $B= - \Umat$ we obtain
\begin{equation}
\det \left[ {\cal Q}^{-1} \cdot M \cdot {\cal P} \right] 
\det[P_+ e^{-\mu L_t} - P_- \Umat]
= 
\det \left[ e^{-\mu L_t} + T \cdot \Umat\right] \,.
\end{equation}
This is the determinant of the reduced Wilson fermion matrix. Note
that the gauge field dependence resides in $T \cdot \Umat$ only and is
completely separated from the dependence on $\mu$. This allows now for
an efficient calculation of the determinant as a function of $\mu$ for
a fixed gauge field background. Denoting the eigenvalues of $T \cdot
\Umat$ by $\lambda_i, i=1,\ldots,4 N_c L_s^3$ we have
\begin{equation}
\det \left[ e^{-\mu L_t} + T \cdot \Umat\right] = \prod_{i=1}^{4 N_c
  L_s^3} (e^{-\mu L_t} + \lambda_i).
\label{eq:reduced matrix}
\end{equation}

In order to establish the equivalence between $\det[M]$ and
eq.(\ref{eq:reduced matrix}) we need to cancel the contribution from
multiplying with $\det[P_+ e^{-\mu L_t} - P_- \Umat]$. First we note
that the matrix has an inverse,
\begin{equation}
(P_+ e^{+\mu L_t} - P_- \Umat^\dagger) \cdot (P_+ e^{-\mu L_t} - P_- \Umat) = 1,
\end{equation}
hence the canceling is always possible. In fact we can calculate the
determinant explicitly, since the matrix splits into the two
orthogonal blocks $\propto P_\pm$ (this is due to the fact that
$\Umat$ and $e^{-\mu L_t}$ are trivial in Dirac space) and the
determinant is just the product of the determinants of the two
subblocks,
\begin{equation}
\det[P_+ e^{-\mu L_t} - P_- \Umat] = \det[e^{-\mu  L_t} ] \cdot \det[-\Umat]= e^{-\mu L_t \cdot 2 N_c L_s^3}
\end{equation}
where the factor of $2$ in the exponent comes from the fact that the
subblock matrix spans over only half the Dirac indices.

\subsection{Calculation of $T$}\label{sec:calculationofT}
A drawback of the matrix reduction is that we need to explicitly
calculate $T$ which contains the inverses of $Q_i^-$ of size $(4 N_c
L_s^3)\times(4 N_c L_s^3)$. It turns out, however, that the size of
the matrix to be inverted can be halved. To see this consider the
following explicit form of $Q^-_i$. The spatial Wilson Dirac operator
$B_i$ on time slice $i$ inherits the structure of the full spatial
Wilson Dirac operator $B$ according to eq.(\ref{eq:spatial_B}). When
the $\sigma_k$ in $\Gamma_k$ are chosen to be hermitian it can be
written as
\be
B_i =  \left( 
\begin{array}{cc}
D_i & C_i \\
-C_i & D_i
\end{array}
 \right)
\ee
with $D_i^\dagger = D_i$ and $C_i^\dagger = C_i$ \footnote{Note that a
  similar argument goes through for the choice of antihermitian Dirac
  matrices in which case one has $C_i^\dagger = -C_i$.}. 
Then, in a basis where $\Gamma_4$ is block
diagonal, one has
\begin{equation}
Q_i^- = P_- + B_i P_+ = \left(
\begin{array}{cc}
D_i & 0 \\
-C_i & 1
\label{eq:Q-}
\end{array}
\right) \, .
\end{equation}
We then find for
the inverse of $Q^-_i$
\begin{equation}\label{eq:Q-_inv}
(Q_i^-)^{-1} = \left(
\begin{array}{cc}
D_i^{-1}              & 0\\
C_i \cdot D_i^{-1}  & 1
\end{array}
\right),
\end{equation}
so $B_i$ needs to be inverted only in the subspace proportional to
$P_+$, i.e.~only $D_i^{-1}$ is needed. Further we also need to
calculate $\det[Q_i^-]$ although these factors cancel exactly in the
ratio of determinants, e.g.~for different chemical potentials. From
eq.(\ref{eq:Q-}) we read off $\det[Q_i^-] = \det[D_i]$.
Finally, for later use we also note the explicit form of $Q_i^+$,
\begin{equation}
Q_i^+ = P_+ + B_i P_- = \left(
\begin{array}{cc}
1 & C_i \\
0 & D_i
\label{eq:Q+}
\end{array}
\right) \, .
\end{equation}

\subsection{Physical interpretation of the reduced matrix}
\label{sec:physical_interpretation}
It is straightforward to give a physical interpretation of the
dimensionally reduced Wilson Dirac fermion matrix. The equivalence of
the determinants
\be\label{eq:reduced_determinant2}
\det M = \det{\cal Q} \cdot  
\det \left[ e^{-\mu L_t/2} + T\cdot \Umat \cdot e^{+\mu L_t/2}\right]
\ee
establishes an equivalence (up to the bulk term $\det{\cal Q}$)
between the four-dimensional Wilson Dirac operator $M$ and the
effective three-dimensional operator $e^{-\mu L_t/2} + T\cdot \Umat
\cdot e^{+\mu L_t/2}$, similar to the equivalence between the
five-dimensional domain wall fermion operator and the four-dimensional
overlap Dirac operator \cite{Borici:2004pn,Edwards:2005an,Edwards:2010}. The analogy
becomes even more transparent at $\mu=0$ when the reduced operator
becomes $1+T\cdot \Umat$, similar to the massless overlap Dirac
operator. The factors $e^{\pm \mu L_t/2}$ can then be understood as
mass terms for the fermions and anti-fermions propagating forward and
backward in time.

This picture is corroborated by the matrix $\tilde M$ in
eq.(\ref{eq:transfer_matrix}) which is essentially a transfer matrix
describing fermions and (anti-)fermions hopping forward and backward
between time slices, respectively. The dynamics of these hoppings are
described by the transfer matrices $T_i$ which, however, are
independent of $\mu$.  Fermions winding around the time direction in
forward direction will eventually pick up a factor $e^{+\mu L_t}$
(residing in $A^+$) for each winding while the anti-fermions winding
around the time direction in backward direction will pick up
corresponding factors of $e^{-\mu L_t}$ (residing in $A^-$). In
addition, the winding fermion modes are then weighted by $\Umat$ which
contains the temporal gauge field dynamics. In that sense, the reduced
matrix is equivalent to a fermion winding number expansion.

To complement this interpretation it is worthwile to consider
alternative forms of the reduced matrix. In the present form the reduced
matrix $e^{-\mu L_t/2} + T\cdot \Umat \cdot e^{+\mu L_t/2}$ makes the
propagation of the fermion forward in time explicit. One can equally
well emphasize the propagation of the anti-fermion backwards in
time. This can for example be achieved by considering, instead of
${\cal Q}^{-1}\cdot M \cdot {\cal P}$, the dimensional reduction of
${\cal P} \cdot M \cdot {\cal \tilde Q}^{-1}$ where ${\cal \tilde Q}$
is the block diagonal matrix containing $Q_i^+$ along the
diagonal. The reduced matrix then becomes
\be
e^{+\mu L_t/2} + \Umat^\dagger \cdot\tilde T \cdot e^{-\mu L_t/2}
\ee
where 
\be
\tilde T = \tilde T_{L_t-1} \cdot \ldots \cdot \tilde T_1 \cdot
\tilde T_0 \quad \text{with } \quad \tilde T_i = T_i^{-1} \, ,
\ee
so making the backward propagation of the anti-fermions (and their
weighting with $-\mu$ instead of $+\mu$) explicit.
\footnote{We note that further equivalent variants of the reduced
matrix can be obtained. By considering projection of $M$ with ${\cal P}^\dagger$ instead of ${\cal P}$ from
the left or right, one
is lead to reduced matrices with modified $T'$ or $\tilde T' =
T'^{-1}$ related to the original $T$ by $T' = \tilde
T^\dagger$.}

Finally, the construction of the reduced matrix presented in section
\ref{sec:detM_reduction} is done for gauge field configurations fixed
to temporal gauge. However, the construction can easily be extended to
the generic case without gauge fixing, leading to a reduced matrix
where the structure $T\cdot \Umat$ becomes
\be
\label{eq:TU_generic}
T_0 \cdot \Umat_0 \cdot T_1 \cdot \Umat_1 \cdot \ldots \cdot T_{L_t-1}
\cdot \Umat_{L_t-1} = \prod_{i=0}^{L_t-1} T_i \cdot \Umat_i \, .
\ee
Here the matrices $\Umat_t$ now collect all the temporal gauge links
at fixed time coordinate $t$, so the matrix $\Umat_{L_t-1}$ is just
$\Umat$ from before.  The factors in the product of
eq.(\ref{eq:TU_generic}) can be cyclically permuted without changing
the physical content, i.e.~the spectrum of the reduced matrix. This is
due to the fact that all the cyclic permutations are related to each
other by similarity transformations (involving the matrices $\Umat_i$
and $T_i$ which have determinant one, cf.~section
\ref{sec:TU_symmetry}) while the first term in the reduced matrix is
trivial in Dirac and color space.

\section{Spectral properties of the reduced matrix}
\label{sec:spectral_properties}

\subsection{Symmetry of  $T \cdot \Umat$}\label{sec:TU_symmetry}
In the construction of the reduced Wilson fermion matrix the crucial
object is the matrix $T \cdot \Umat$. It turns out that this matrix
has interesting properties which express themselves in peculiar
symmetry properties of the eigenvalue spectrum.

The first thing to note is that 
\be
\det T \cdot \Umat = 1 \,.
\ee
This can easily be seen from eq.(\ref{eq:Q-_inv}) and eq.(\ref{eq:Q+})
where we read off $\det (Q_i^-)^{-1} = \det D_i^{-1}$ and $\det Q_i^+
= \det D_i$, respectively, and hence $\det T_i = \det \left[(Q_i^-)^{-1} \cdot
  Q_i^+ \right]= 1$.

Secondly, we note that the eigenvalues of $T \cdot \Umat$ come in
pairs: for every eigenvalue of $\lambda$, there is an eigenvalue
$\lambda' = 1/\lambda^{*}$. This can be seen as follows. The product
$T_i = (Q_i^-)^{-1} \cdot Q_i^+$ can be LDU-decomposed with the help
of eq.(\ref{eq:Q-_inv}) and eq.(\ref{eq:Q+}):
\be \label{eq:LDU}
 (Q_i^-)^{-1} \cdot Q_i^+ = \left(
\begin{array}{cc}
1 & 0\\
C_i & 1
\end{array}
\right)
\left(
\begin{array}{cc}
D_i^{-1} & 0 \\
0        & D_i
\end{array}
\right)
\left(
\begin{array}{cc}
1 & C_i \\
0 & 1
\end{array}
\right) \,.
\ee
In this form it is easy to calculate its hermitian conjugated inverse, i.e.
\be
\left[ \left( (Q_i^-)^{-1} \cdot Q_i^+ \right)^{-1} \right]^\dagger = 
\left(
\begin{array}{cc}
1   & -C_i \\
0   & 1
\end{array}
\right)
\left(
\begin{array}{cc}
D_i & 0 \\
0   & D_i^{-1}
\end{array}
\right)
\left(
\begin{array}{cc}
1 & 0\\
-C_i  & 1
\end{array}
\right).
\ee
Comparing this with eq.(\ref{eq:LDU}) we find that
\be
\left[ \left( (Q_i^-)^{-1} \cdot Q_i^+ \right)^{-1} \right]^\dagger =
S \cdot \left( (Q_i^-)^{-1} \cdot Q_i^+ \right) \cdot S^{-1} \quad
\text{with } \quad
S = \left(
\begin{array}{cc}
0 & 1\\
-1 & 0
\end{array}
\right),
\ee
and hence
\be
\left[ \left(  T \cdot \Umat \right)^{-1} \right]^\dagger = S
\cdot \left( T \cdot \Umat \right) \cdot  S^{-1} \, .
\ee
As a consequence the matrix $T \cdot \Umat$ shares the eigenvalue
spectrum with its hermitian conjugated inverse, that is, for each
eigenvalue $\lambda \in \text{spec}\, (T \cdot \Umat)$ there is
another eigenvalue $1/\lambda^* \in \text{spec}\, (T \cdot
\Umat)$. The spectral symmetry hints at the possibility that the
reduced matrix could be further compressed in size by a factor of two
without loosing any spectral information. In principle this can be
achieved by the projection of $T \cdot \Umat$ to a suitable subspace,
but so far we have not been able to construct such a projection,
essentially due to the fact that $T \cdot \Umat$ is non-normal.

\subsection{Eigenvalue distribution of  the reduced matrix }
So far we have been concerned with purely algebraic properties of the
Wilson Dirac matrix $M$ and the corresponding reduced matrix $T \cdot
\Umat$. In practice, what is needed are all the eigenvalues
$\lambda_i$ of the latter matrix, such that the determinant of
$M(\mu)$ can be evaluated for any arbitrary value of the chemical
potential according to
\be\label{eq:reduced_determinant}
\det M(\mu) = \det{\cal Q} \cdot  e^{+\mu L_t \cdot 2 N_c L_s^3} \prod_{i=1}^{4
  N_c   L_s^3} (e^{-\mu L_t} + \lambda_i).
\ee
Apart from the symmetry property $\lambda \leftrightarrow
1/\lambda^{*}$ discussed in section \ref{sec:TU_symmetry} the
eigenvalue spectrum has additional interesting features.  To
illustrate these we will use configurations generated for a previous
study of non-zero baryon density systems using the canonical partition
function~\cite{Li:2010qf}. These $6^{3}\times 4$ configurations are
picked from $N_{f}=4$ ensembles at a temperature close to the
deconfining transition: $T\approx 0.95 T_{c}$. The parameters for the
fermionic matrix will be set to the values used to generate the
ensembles: $\kappa = 0.1371$ and $c_\text{sw}=1.96551$ corresponding
to a pion mass of about $700-800\,\text{MeV}$.

\begin{figure}[!t]
   \centering
   \includegraphics[width=13cm]{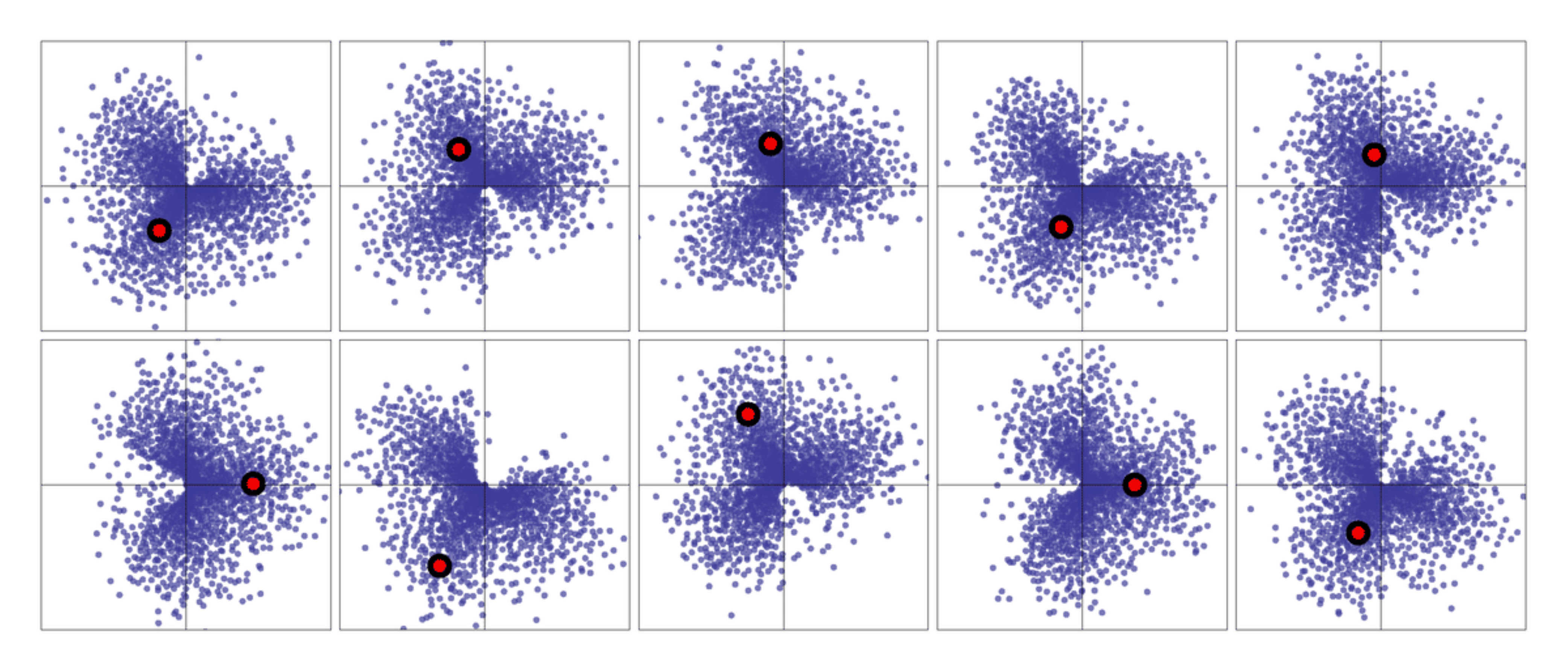}
   
   \includegraphics[width=13cm]{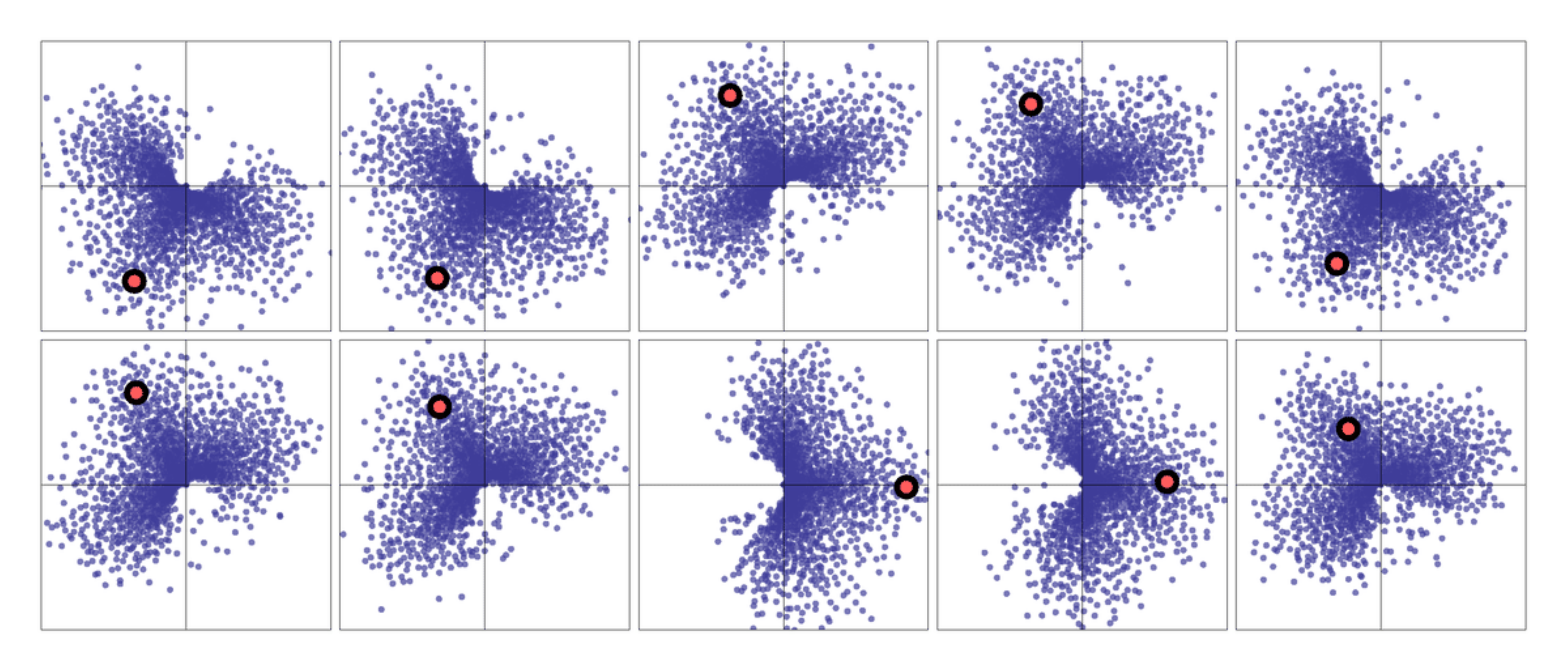}
   \caption{The eigenvalue distribution for 10 arbitrary
     configurations drawn from canonical ensembles with $n_{B}=4$
     (top) and $n_{B}=11$ (bottom). The red point indicates a scaled
     value of the spatially averaged Polyakov loop to show its
     correlation with the eigenvalue distribution.}
   \label{fig:TU_ev}
\end{figure}

In Fig.~\ref{fig:TU_ev} we plot the eigenvalue distribution of the
matrix $T \cdot \Umat$ for a random selection of gauge field
configurations drawn from canonical ensembles with baryon number
$n_B=4$ (top two rows) and $n_B=11$ (bottom two rows), respectively.
The plots represent the eigenvalue distribution in the complex plane
and the scale is from ${\rm Re, Im}\,\lambda \in [-600, 600]$. Only
the large magnitude eigenvalues are visible (the low magnitude members
are also plotted but they are not visible on this scale). Note that
there is a cone empty for each configuration and that the
distributions exhibit a three lob structure which, from configuration
to configuration, is related by $\Z(3)$-rotations in the complex
plane. An interesting observation is the fact that for each
configuration the structure is correlated with the value of the
spatially averaged Polyakov loop, which in the temporal gauge is given
by
\be
P(U) = \frac{1}{4 N_c L_s^3} \, \text{tr } \Umat \, .
\ee
In Fig.~\ref{fig:TU_ev} we make this correlation explicit for each
configuration by imposing the corresponding (rescaled) value of $P(U)$
onto the eigenvalue spectrum. Of course this correlation is no
surprise, since from the structure $T\cdot \Umat$ it is immediately
clear that under a $\Z(3)$-transformation of the temporal gauge fields
$\Umat$ the eigenvalues are simply rotated by the corresponding
$\Z(3)$ factor.  On the other hand, one should also keep in mind that
some of the correlations we observe between the determinant and the
Polyakov loop might be due to the rather heavy quark mass
\cite{deForcrand:1999cy} and the correlations might become less
pronounced as we move towards lower quark masses.

\begin{figure}[!t]
   \centering
  \includegraphics[width=15cm]{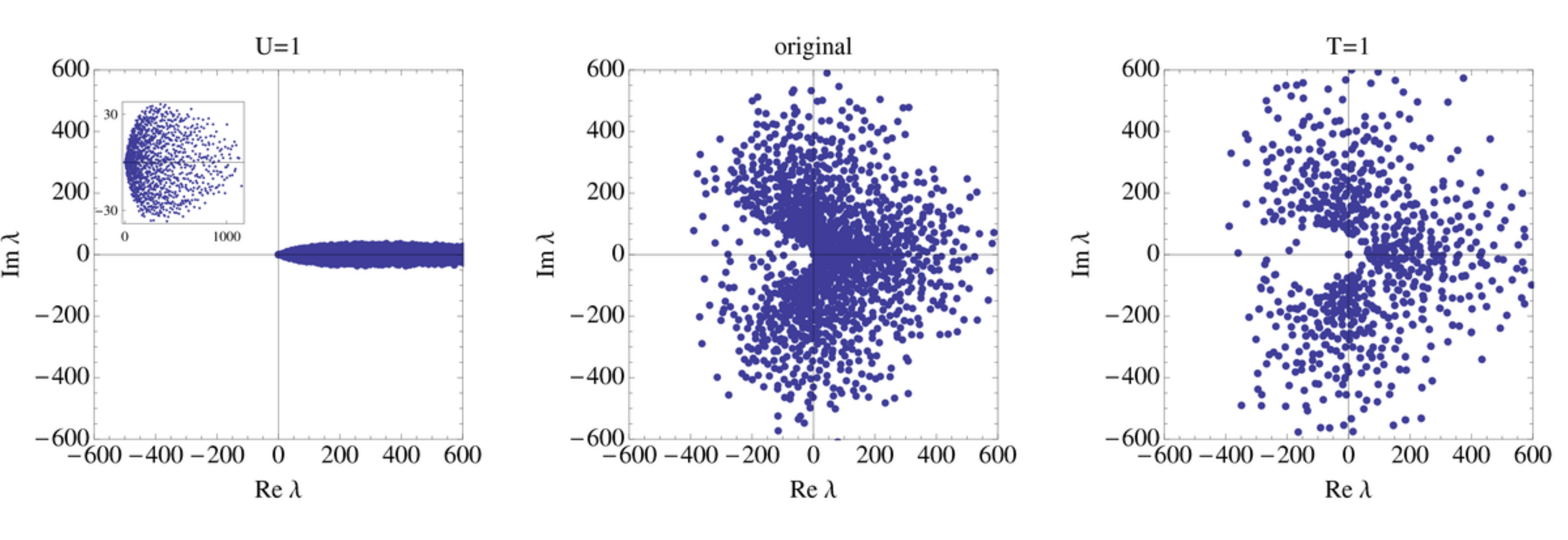}
  \caption{The eigenvalue spectrum of $T \cdot \Umat$ for an
    arbitrary configuration with $\arg P(U) \approx 0$ (middle plot), in the case
    when $\Umat = 1$ is put by hand (left plot), or when all spatial
    links are set to one (right plot).}
   \label{fig:modified_TU}
\end{figure}

\begin{figure}[!t]
   \centering
  \includegraphics[width=15cm]{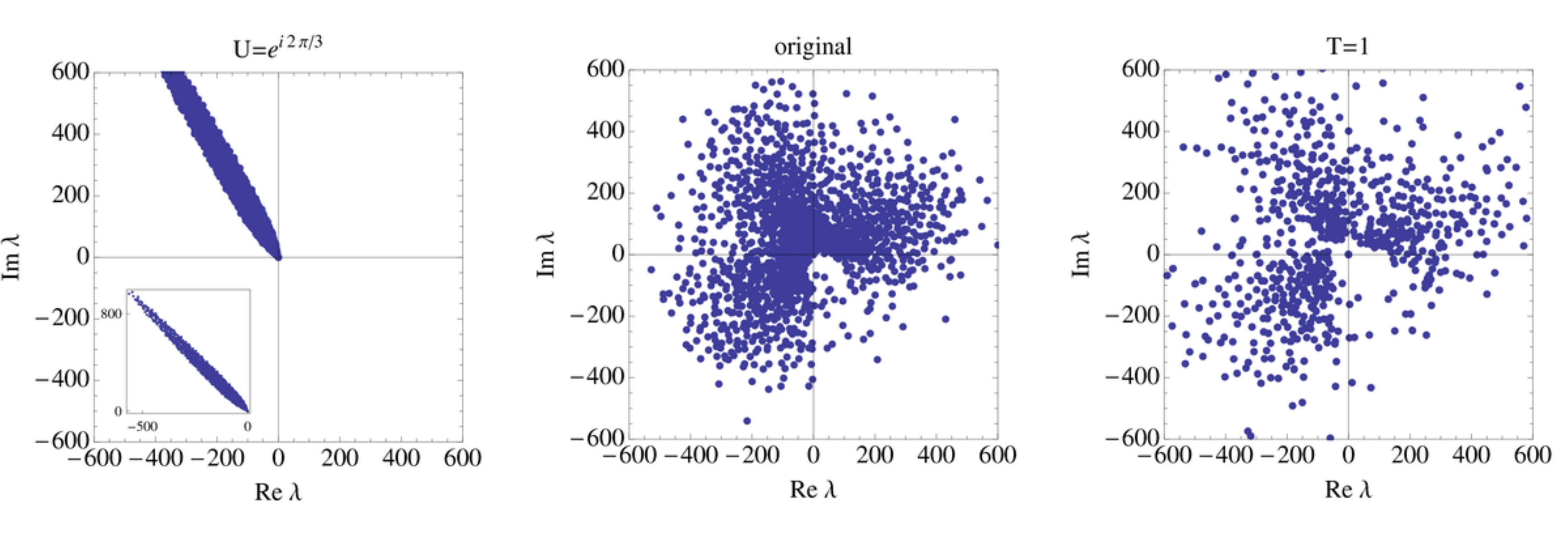}
  \includegraphics[width=15cm]{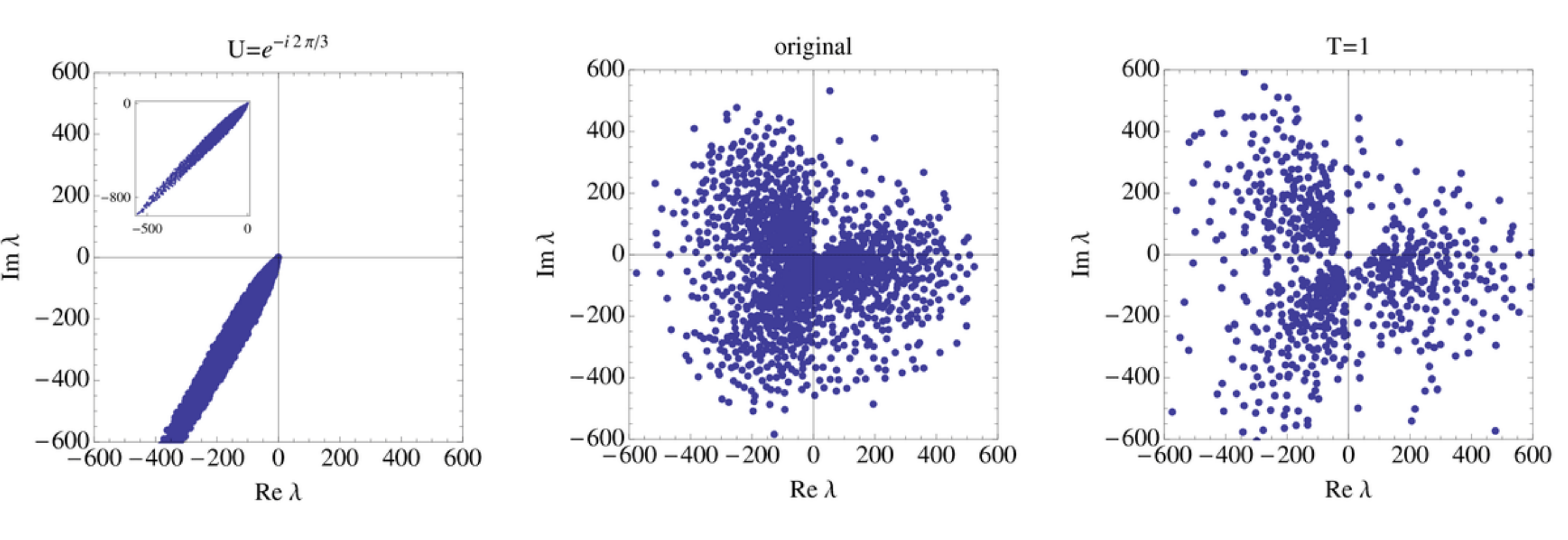}
  \caption{Same as figure \ref{fig:modified_TU}, but for a configuration
    with $\arg P(U) \approx + 2 \pi/3$ (top row) and one with $\arg P(U)
    \approx - 2 \pi/3$.}
   \label{fig:modified_TUZ3}
\end{figure}

In order to further expose the influence of the $\Z(3)$ phase of
$P(U)$ (or rather $\Umat$) on the spectrum, we perform the following
exercise. Instead of computing the eigenvalues of the original reduced
matrix $T \cdot \Umat$, we calculate the spectrum of a modified
reduced matrix where we set the temporal gauge fields $\Umat$ to the
$\Z(3)$ phase as given by the Polyakov loop $P(U)$. We show the result
of this calculation for a configuration with $\arg P(U) \simeq 0$ in
Fig.~\ref{fig:modified_TU} in the left most plot. The only gauge field
dependence is now through the spatial gauge links in $T$ and we see
that this dependence is responsible for the eigenvalues' variation in
magnitude, while the phase of the eigenvalues fluctuates very weakly
around zero. (Note that the spectrum of the reduced free Wilson Dirac
operator is real and the eigenvalues $\lambda>1$ span the range
between roughly 6 and 2000.)
Of course we can also turn the argument around and put all the spatial
links to unity, so that the gauge field dependence resides in $\Umat$
alone. The spectrum of the reduced matrix modified in this way is
shown in the right most plot in Fig.~\ref{fig:modified_TU}. Comparing
this with the original spectrum shown in the middle plot we conclude
that the phase variation of the eigenvalues is determined almost
solely by the temporal gauge fields $\Umat$ while the spatial gauge
fields only add small fluctuations to the phase. In
Fig.~\ref{fig:modified_TUZ3} we repeat this exercise for
configurations with $\arg P(U) \simeq \pm 2\pi/3$.  As discussed
above, in this case the spectra are simply rotated by the
corresponding $\Z(3)$ factors, and the conclusion remains the same.

\section{Projected determinant of the canonical partition function}
\label{sec:ReductionResults}
After having discussed the properties of the reduced matrix and its
spectrum, the next interesting quantity to study is the determinant as
a function of the (real or imaginary) chemical potential, and
eventually also the projected determinant which subsumes the dynamics
of the fermions in the canonical partition function.

\subsection{The determinant at non-zero chemical potential}
\begin{figure}[!t]
   \centering
   \includegraphics[width=13cm]{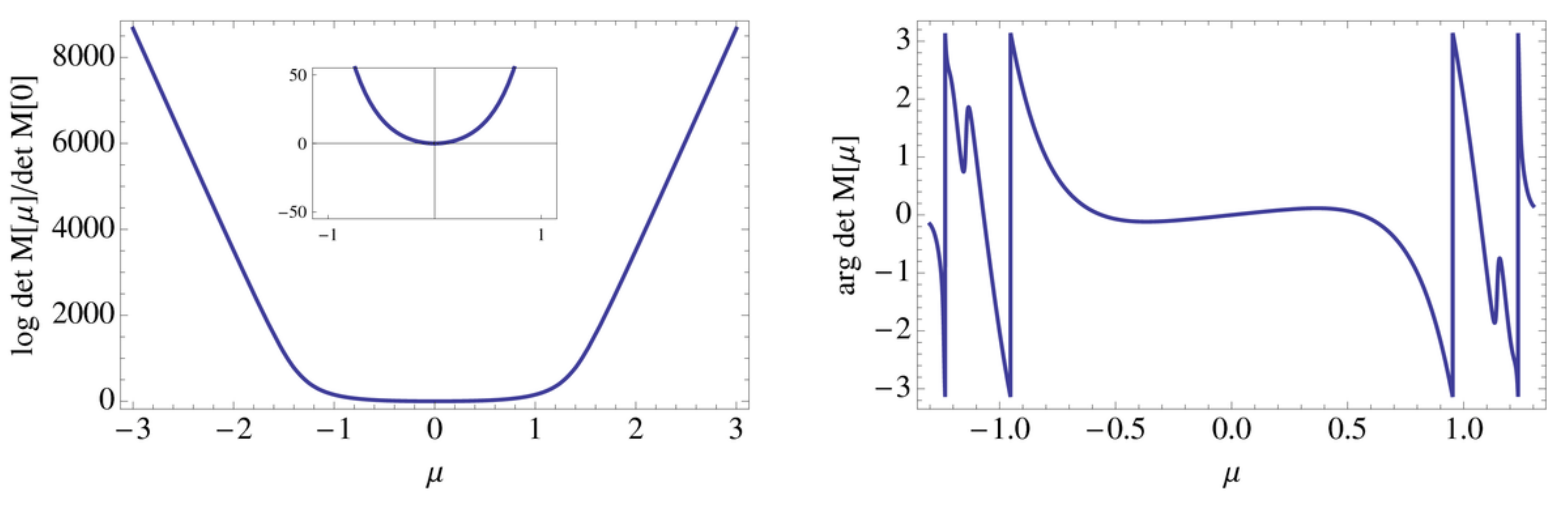}
   \includegraphics[width=13cm]{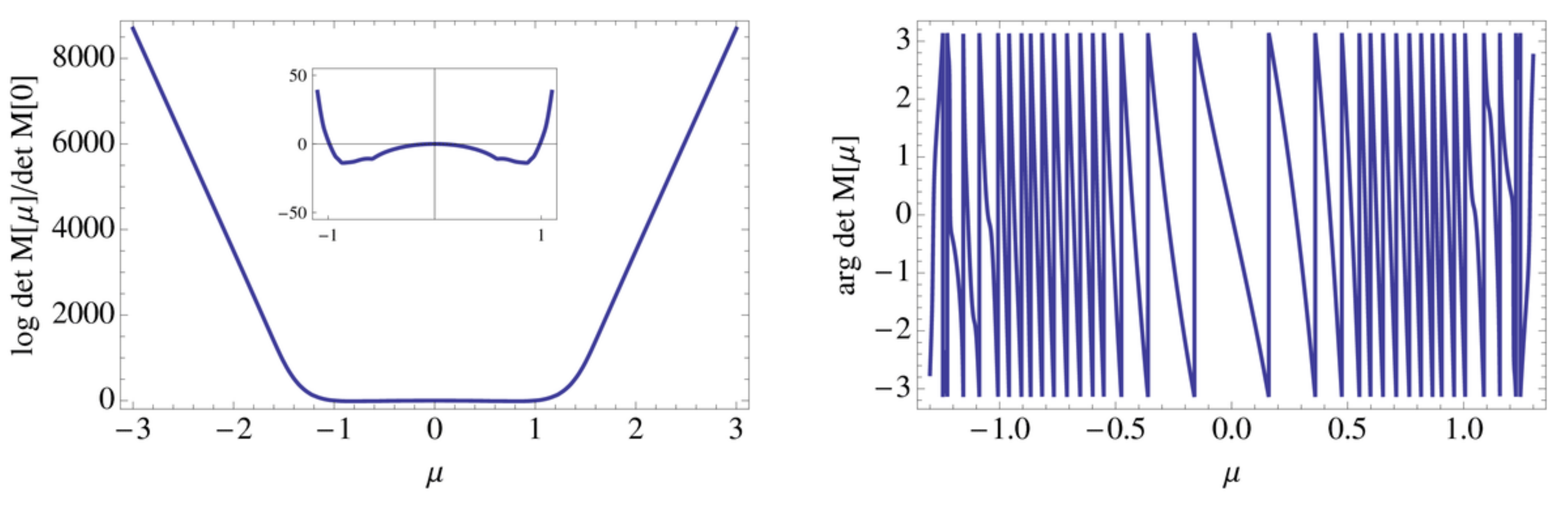}
   \includegraphics[width=13cm]{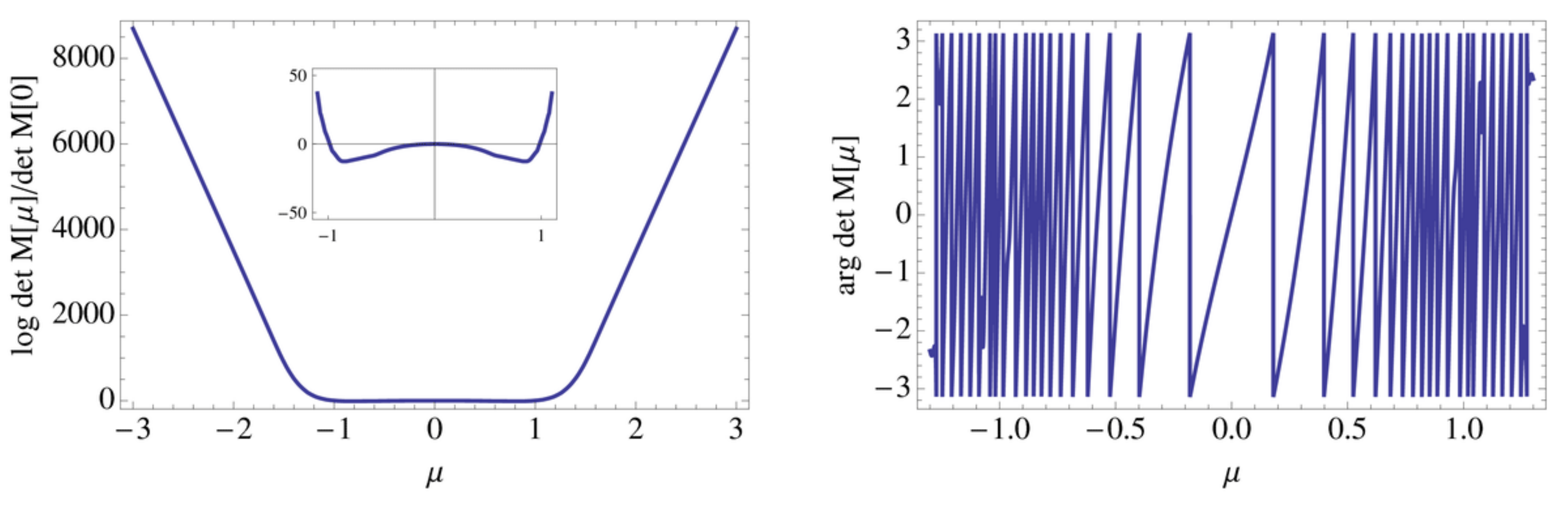}
   \caption{The logarithm of the determinant $|\det M(\mu)|$ normalised
     to $\det M(\mu=0)$ (left row) and the argument of $\det M(\mu)$
     (right row) for three configurations from a canonical ensemble
     with $n_B=4$. From top to bottom the configurations have $\arg
     P(U) \approx 0,\, +2\pi/3$ and $-2 \pi/3$.}
   \label{fig:detmu_k12}
\end{figure}

Having the eigenvalues $\lambda_i$ of $T\cdot \Umat$ at hand it is now
easy to evaluate the determinant of the Wilson Dirac operator for
arbitrary chemical potential according to
eq.(\ref{eq:reduced_determinant}). In Fig.~\ref{fig:detmu_k12} we show
the logarithm of the determinant $\det M(\mu)$ for three
configurations from a canonical ensemble with $n_B=4$. From top to
bottom the configurations have $\arg P(U) \approx 0,\, +2\pi/3$ and
$-2 \pi/3$.
The first row shows the logarithm of $|\det M(\mu)|$ normalized to
$\det M(\mu=0)$, i.e.~$|\log \det M(\mu)|/\det M(\mu=0)$, while the
second row shows the argument $\arg \det M(\mu)$ modulo $2\pi$. We
note significant differences between the results for configurations in
the various triality sectors. Firstly, while in all sectors the size
of the determinant ratio starts to grow exponentially with $|\mu|$
beyond $|\mu| \gtrsim 1$, configurations in the non-trivial
$\Z(3)$-sectors show a local minimum just below $|\mu| \sim 1$ in
contrast to configurations in the trivial sector which show a
monotonic increase with $|\mu|$. Secondly, the derivative of the phase
w.r.t.~$\mu$ stays roughly constant for $|\mu| \lesssim 0.5$, but
depends strongly on the $\Z(3)$-sector.

\begin{figure}[!t]
   \centering
   \includegraphics[width=13cm]{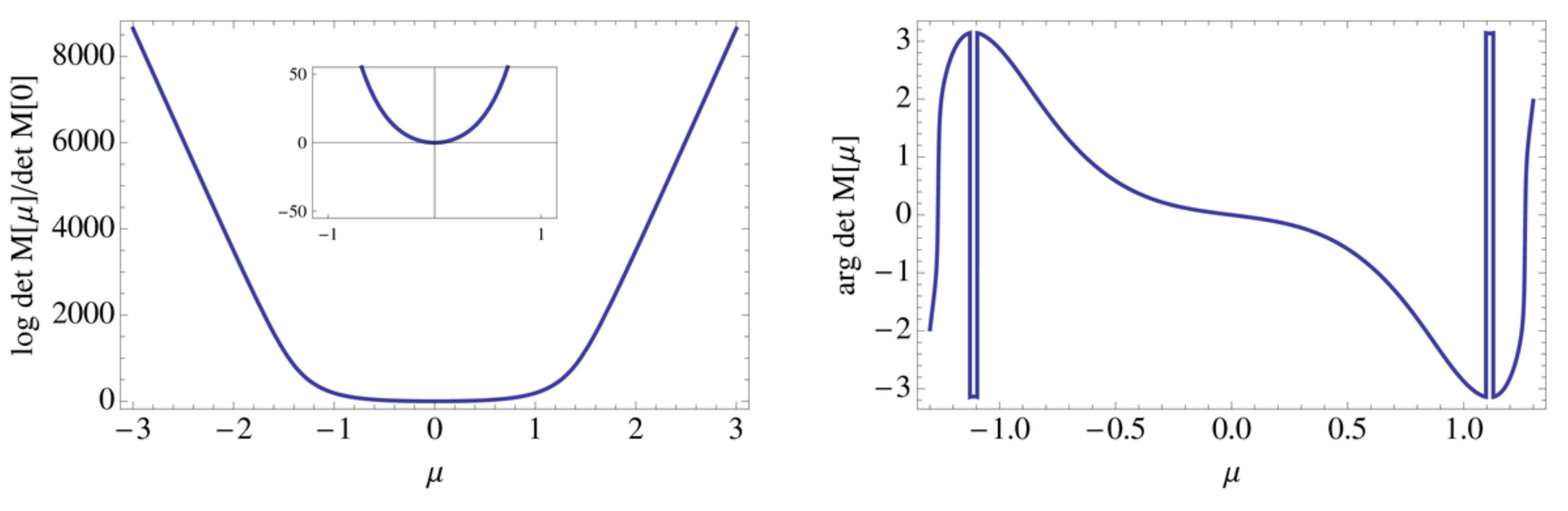}
   \includegraphics[width=13cm]{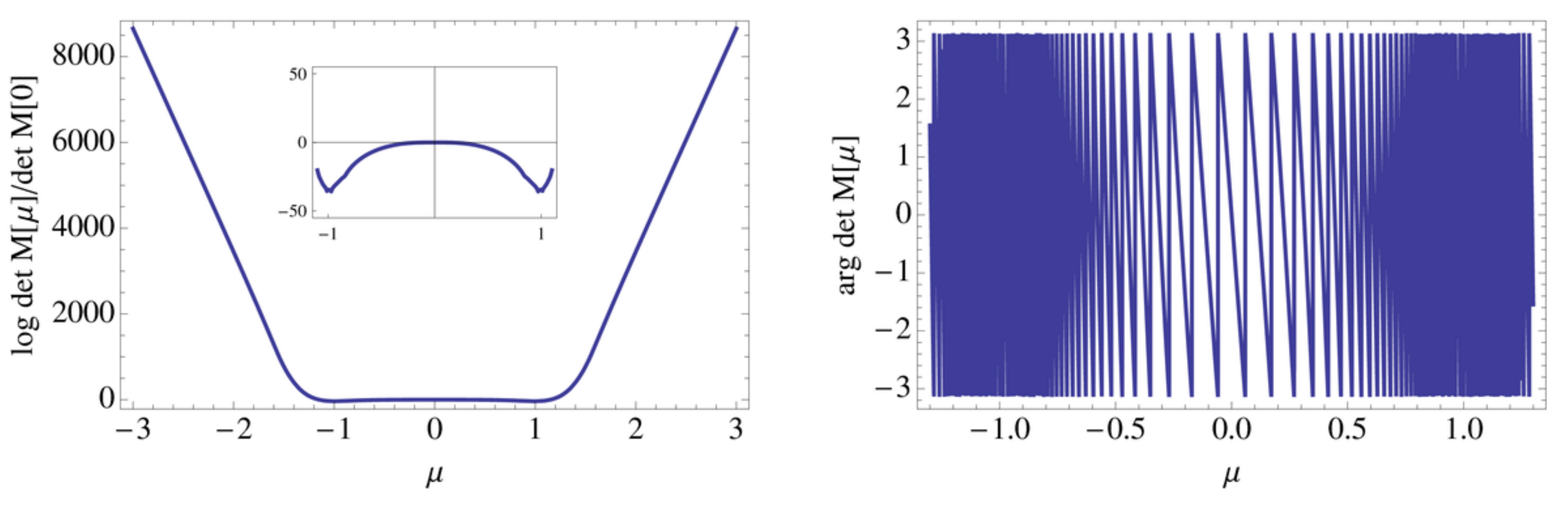}
   \includegraphics[width=13cm]{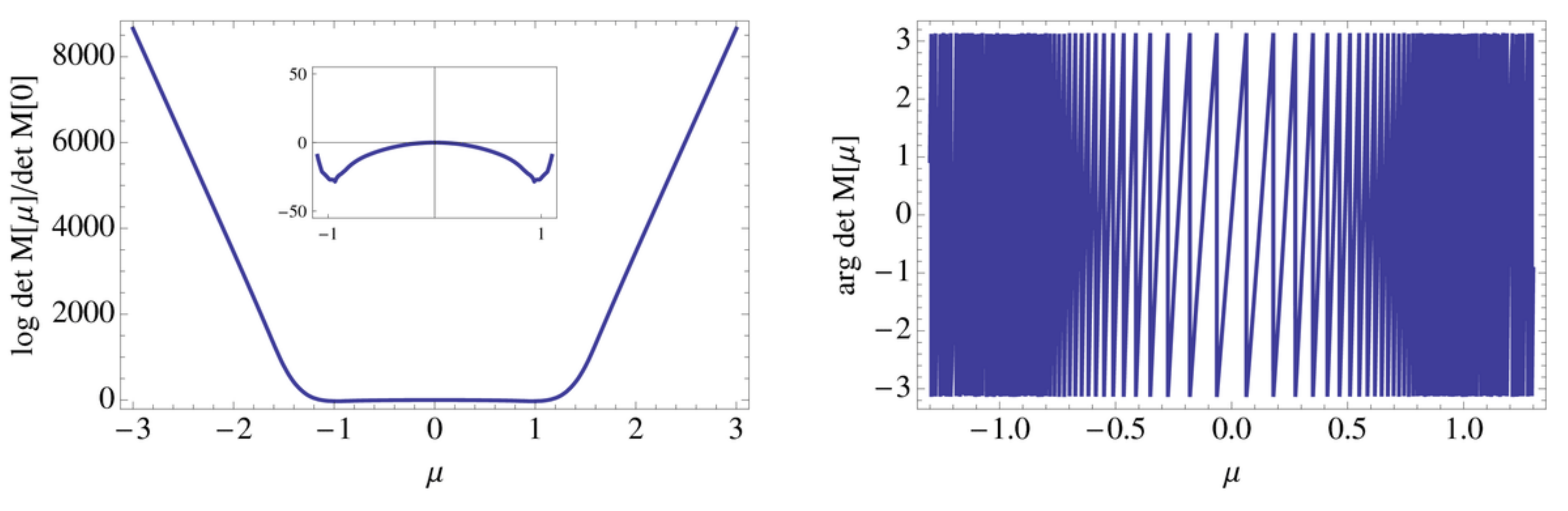}
   \caption{Same as Fig.~\ref{fig:detmu_k12}, but for three
     configurations from a canonical ensemble with $n_B=11$.}
  \label{fig:detmu_k33}
\end{figure}
In Fig.~\ref{fig:detmu_k33} we show the same kind of plots for three
typical configurations (with $\arg P(U) \approx 0,\, +2\pi/3$ and $-2
\pi/3$ from top to bottom) from a canonical ensemble with
$n_B=11$. The picture qualitatively remains the same except that the
features described before are accentuated. For configurations in the
non-trivial $\Z(3)$-sectors the minima of $|\det M(\mu)|$ become
slightly deeper and move to slightly larger values of
$|\mu|$. Furthermore, for those configurations the phase of $\det
M(\mu)$ changes more rapidly while the change of the phase in the
trivial $\Z(3)$-sector becomes smoother and stretches further into
larger values of $|\mu|$, possibly allowing reweighting to larger
chemical potential.

\begin{figure}[!t]
   \centering
   \includegraphics[width=13cm]{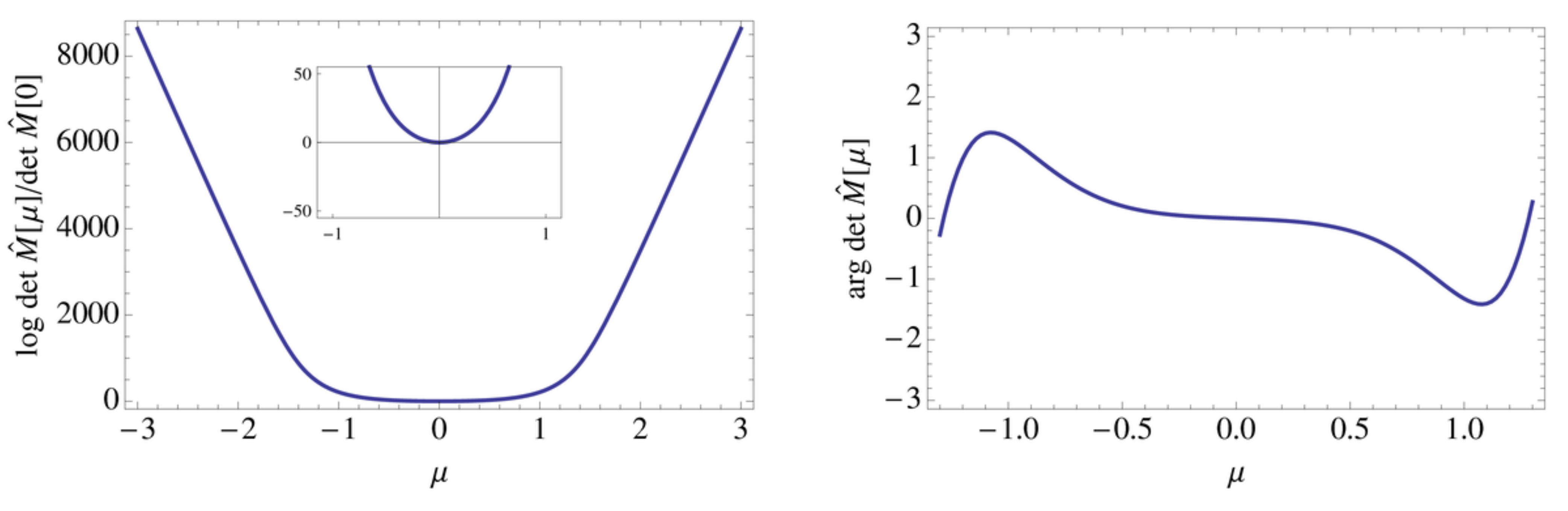}
 \caption{Same as the middle panel of Fig.~\ref{fig:detmu_k33} but for
 the modified determinant eq.(\ref{eq:modified_determinant}).}
   \label{fig:detmu_k33_mod}
\end{figure}

Note that the wild phase fluctuations observed for configurations with
non-trivial Polyakov loop are due to the contributions from the
fractional baryon number sectors. To illustrate this point, following
the ideas presented in~\cite{Kratochvila:2006jx}, we define a modified
determinant that only includes the integral baryon number sectors:
\begin{equation}\label{eq:modified_determinant}
\det \hat{M}(U, \mu) \equiv \sum_{n_{B}} e^{3 n_{B} \mu/T} {\det}_{3 n_{B}} M(U).
\end{equation}
This definition is equivalent to the fugacity expansion for the
determinant where we only sum over the sectors that have a net number
of quarks divisible by 3.  In figure~\ref{fig:detmu_k33_mod} we plot
the absolute value and phase of this modified determinant for a
configuration with $\arg P(U)\approx 2\pi/3$ (the same one used in the
middle panel of figure~\ref{fig:detmu_k33}). We see that the magnitude
now increases monotonically and that the phase changes much slower
with $\mu$ -- this is exactly the behavior of configurations that have
$\arg P(U)\approx 0$.

As we emphasized in the introduction these observations are hard to
interpret physically, but we believe that they might play an important
role for the optimization of reweighting strategies, or for the
development of new canonical simulation algorithms.

\subsection{Calculation of the projected determinants}

Using eq.(\ref{eq:reduced_determinant}) we can show that the projected
determinants $\det_k M$ defined in eq.(\ref{eq:def_detk}), up to a multiplication
with $\det {\cal Q}$, are the coefficients $c_{k+k_{\rm max}}$ of the
polynomial
\begin{equation}
\Pi(x) = \prod_{i=1}^{2 k_{\rm max}} (x + \lambda_{i}) = \sum_{k=0}^{2 k_{\rm max}} c_{k} x^{k},
\end{equation}
where $k_{\rm max}= 2N_{c}N_{s}^{3}$.  A couple of coefficients can be
computed easily: $c_{2k_{\rm max}} = 1$ and $c_{0} = \prod_i \lambda_i
= \det T \cdot \Umat = 1$ as discussed before. All other coefficients can be
calculated recursively. To show this, we first define the partial
product:
\begin{equation}
\Pi_{n}(x) = \prod_{i\leq n}(x+\lambda_{i}) = \sum_{k\leq n} c_{k}^{(n)} x^{k}.
\end{equation}
Clearly, $\Pi_{n+1}(x) = (x+\lambda_{n+1}) \Pi_{n}(x)$ and hence we have
\begin{equation}
c^{(n+1)}_{k}= \lambda_{n+1} c^{(n)}_{k} + c^{(n)}_{k-1},
\label{eq:iteration}
\end{equation}
for all $0\leq k \leq n+1$ (we set $c^{(n)}_{-1}=0$). For $\Pi_{0}$
all coefficients are zero except for $c^{(0)}_{0} = 1$. Using
eq.(\ref{eq:iteration}) to compute $c^{(n+1)}$ from $c^{(n)}$, after $2
k_{\rm max}$ steps we obtain the coefficients of $\Pi_{2 k_{\rm max}}
\equiv \Pi$.

The recursive steps of the iteration have to be carried out using a
high precision library. The reason for this is that the magnitude of
the coefficients vary over thousands of orders of magnitude. We used
GNU Multi-Precision library which can easily handle numbers of this
magnitude. While the calculation takes significantly longer than when
using the standard floating point arithmetic, the total time is small compared to
the time it takes to compute the eigenvalues of the reduced
matrix. One possible issue when using a high precision library is that
the results look deceptively precise since the inputs are treated as
high precision numbers when their real precision is at the level of
machine precision or less. To check the robustness of our calculation
we performed the following tests. In the first test we added random
numbers of the order of $10^{-15}$ (double precision level) to the
links and recomputed the projected determinants; the relative change
in the projected determinants was at the level of $10^{-10}$. Next, we
randomly reordered the eigenvalues of the reduced matrix and repeated
the recursive step. We find that the relative change in the
coefficients is of the order of $10^{-9}$. We conclude that our
procedure is robust and it produces accurate results.

Before we conclude this section, we want to point out that the
$\lambda\leftrightarrow 1/\lambda^{*}$ symmetry of the reduced matrix,
together with the fact that $\det T \cdot \Umat = 1$ can be used to show that
$c_{k_{\rm max}+k} = c_{k_{\rm max}-k}^{*}$. Since $\det {\cal Q}$ is
real, this insures that that ${\det}_{k} M = ({\det}_{-k} M)^{*}$ a
fact easily derived from the definition of the projected determinant
and the reality of $\det M(\phi)$.

\subsection{Size distribution of the projected determinant}

\begin{figure}[!t]
   \centering
  \includegraphics[width=13cm]{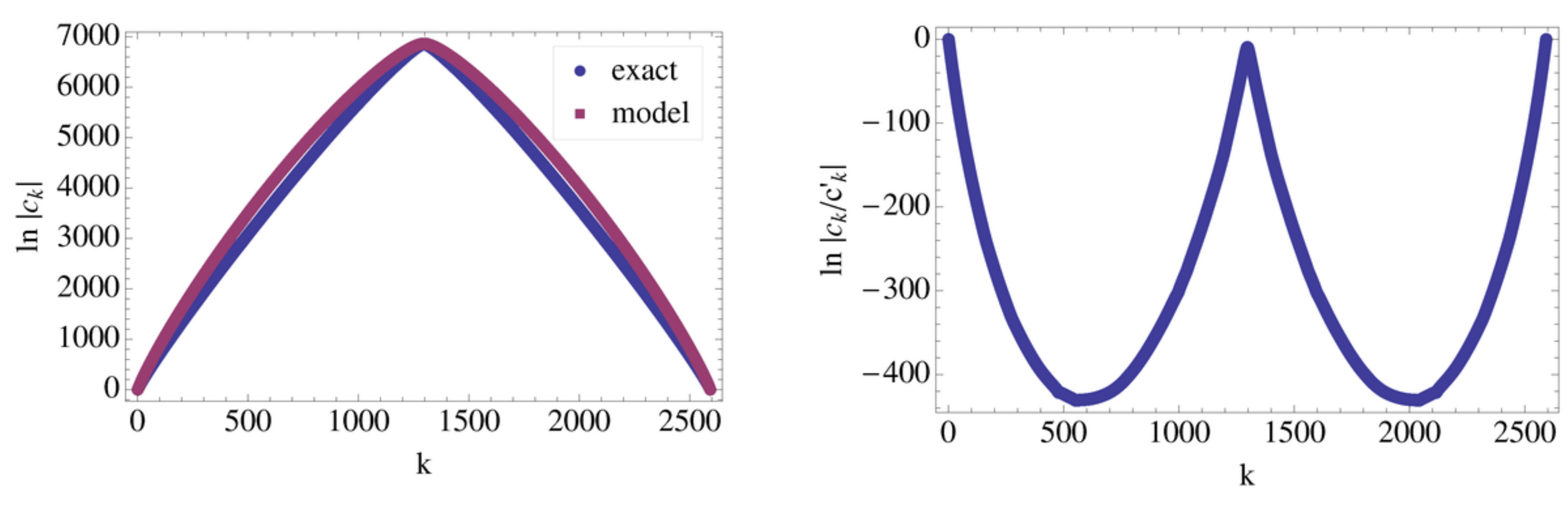}
  \caption{Simple model calculation: on the left the exact values and
    the model values are shown. In the right panel the ratio between
    the exact and model values are shown.}
   \label{fig:proj-model1}
\end{figure}
One of the interesting aspects of this calculation is the fact that
the magnitude of the projected determinants varies over many orders of
magnitude as we change the quark number sector. In this section we
will show that the bulk of this variance can be captured by a simple
combinatorial argument.  However, this only captures part of the
variance: in order to describe the variance accurately we need to take
into account the complex phase of the eigenvalues of the reduced
matrix.

In the main, the bulk of the magnitude is given by the average
magnitude of the eigenvalues and combinatorics. To illustrate this
point, in Fig.~\ref{fig:proj-model1} we compare the value of the
coefficients of the polynomial $\Pi(x)$ for a given configuration with
the coefficients of the polynomial $\Pi'(x) =
(x+\bar{\lambda})^{k_{\rm max}} (x+1/\bar{\lambda})^{k_{\rm max}}$,
where $\bar{\lambda}$ is the geometric mean of the magnitude of the
large eigenvalues $\bar{\lambda} = \left( \prod_{|\lambda|>1}
  |\lambda_{i}|\right)^{1/k_{\rm max}}$.

\begin{figure}[!t]
   \centering
  \includegraphics[width=13cm]{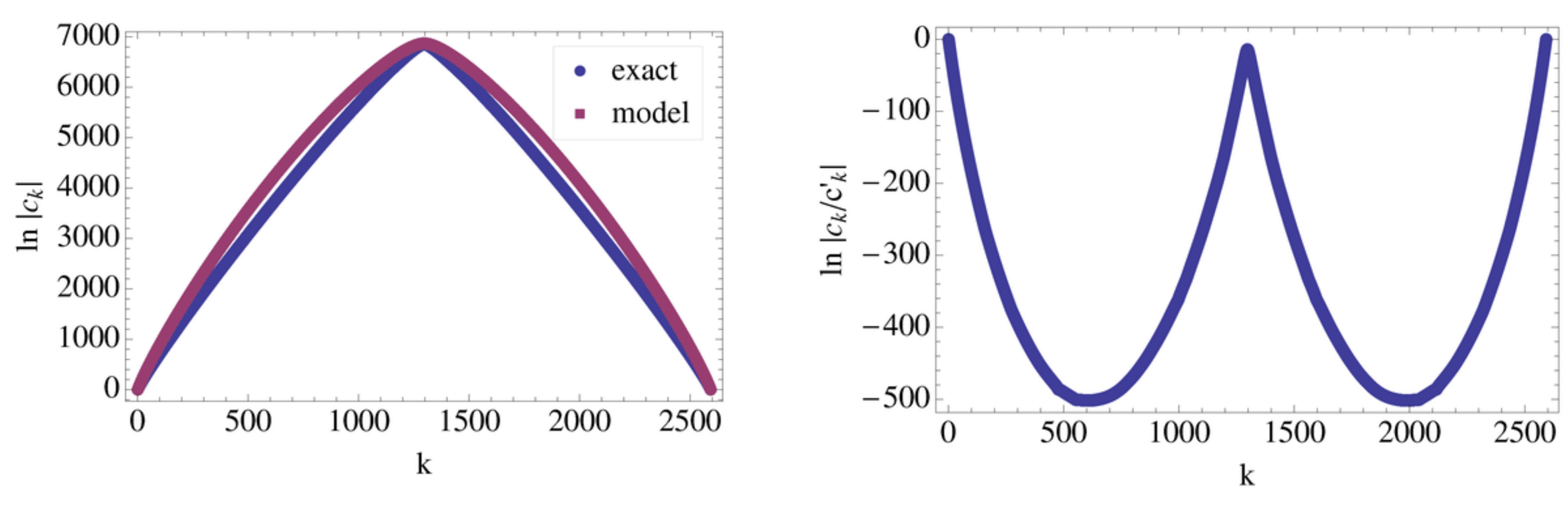}
  \caption{Comparison between true polynomial coefficients and the
    ones where the eigenvalue phase is quenched.}
   \label{fig:proj-model2}
\end{figure}

For any given configurations the eigenvalues vary both in phase and
magnitude; in the comparison above we quenched both
fluctuations. While this approximation captures a good part of the
variation of magnitude, there is still quite a discrepancy left.  To
trace the source of discrepancy we compare the coefficients of
$\Pi(x)$ with a polynomial where each eigenvalue is replaced with its
magnitude: $\Pi''(x) = \prod_{i} (x+|\lambda_{i}|)$. The results of
this comparison are presented in Fig.~\ref{fig:proj-model2}: we see
that the discrepancy is similar to the one above. This shows that the
source of discrepancy is the complex phase fluctuation that is
disregarded here. This conclusion is also supported by the fact that
the coefficients of $\Pi''(x)$ are larger in magnitude than the
coefficients of $\Pi(x)$; this is due to cancelations produced by
phase fluctuations.

\begin{figure}[!thb]
   \centering
  \includegraphics[width=13cm]{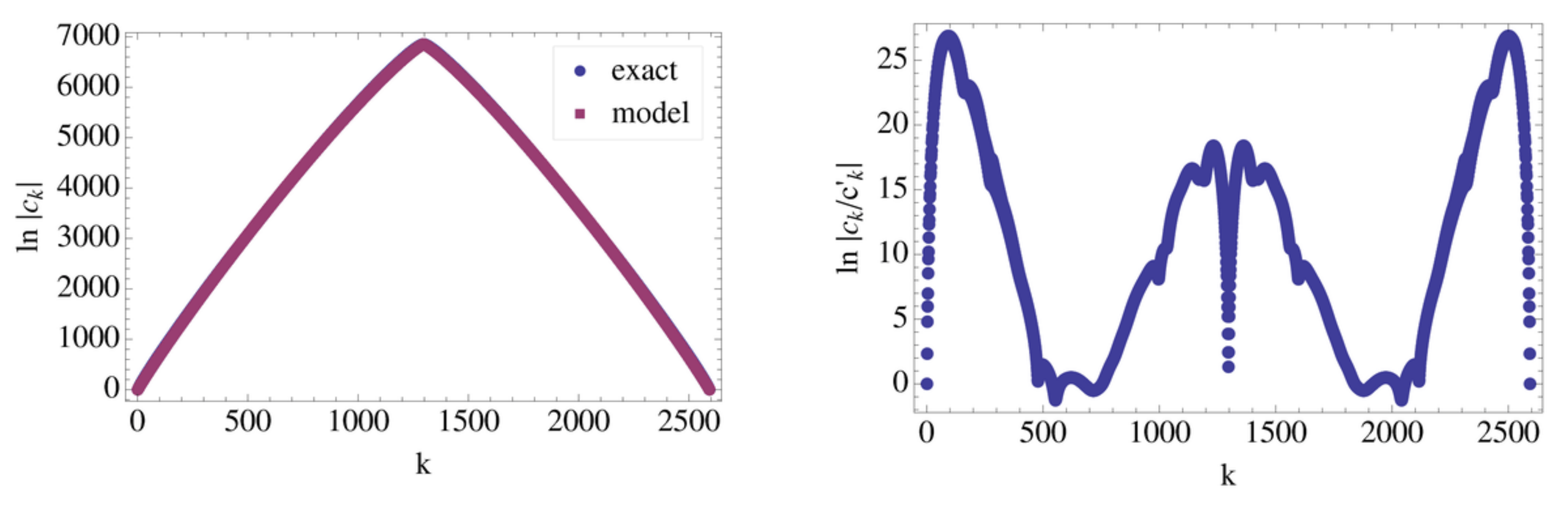}
  \caption{Comparison between true polynomial coefficients and the
    ones where the eigenvalue phase is drawn from a random $\Z(3)$
    distribution. Note the difference in scale in the right plot as
    compared to Fig.~\ref{fig:proj-model2} and \ref{fig:proj-model1}.}
   \label{fig:proj-model3}
\end{figure}

To prove that phase fluctuations are responsible for the discrepancy
we model the rough features of the phase distribution. The phase
distribution is illustrated in Fig.~\ref{fig:TU_ev}. One obvious
feature is that the eigenvalues are concentrated around the $\Z(3)$
axes.  This makes sense: if the eigenvalues were exactly along the
$\Z(3)$ axes, the projected determinants that have zero triality
($k=3n_{B}$) will be real whereas the ones with non-zero triality will
have a fluctuating phase and they will vanish when averaged over gauge
configurations, as expected.

To check that phase fluctuations are responsible for reducing the
magnitude of the coefficients we compare the coefficients of $\Pi(x)$
with $\Pi'''(x)= \prod_{i} (x+ z_{i} |\lambda_{i}|)$, where $z_{i}$ is
a a random $\Z(3)$ phase with one constraint: we enforce the
$(\lambda, 1/\lambda^{*})$ pairing. The results are presented in
figure~\ref{fig:proj-model3}: we see that the results agree much
better now. It is clear that the phase fluctuations play an important
role in determining the coefficients. Phase fluctuations reduce the
value of the coefficients significantly: from
figure~\ref{fig:proj-model2} we see that for intermediate $k$ values
they reduce the magnitude by about 200 orders of magnitude. From
figure~\ref{fig:proj-model3} we see that this is mostly due to the
$\Z(3)$ nature of the fluctuations (we also tested random $\Z(n)$
phases with $n\in\{2,4,5,6\}$ but they fail to produce the same
agreement).

\section{Phase fluctuations in canonical ensembles}\label{sec:canonicalResults}
The reduced Wilson fermion matrix constructed and discussed in the
previous sections allows for several intriguing
applications. Presumably among the most interesting are the direct
simulation of canonical ensembles and the reweighting to different
fermion numbers, or values of the chemical potential. In the following
we briefly discuss these two applications and present some results on
the phase fluctuations encountered in direct simulations of canonical
ensembles, merely to illustrate the capabilities of the reduced
fermion matrix approach.

In order to simulate the canonical partition function
eq.(\ref{eq:def_ZC}) by Monte-Carlo techniques one needs the integrand
to be real and positive. Since in general this is not guaranteed, the
approach so far \cite{Alexandru:2005ix} has been to ensure positivity
by fiat, i.e.~to generate an ensemble using the weight $W_{|k|}(U)
\propto |{\rm Re}\,{\det}_{k} M(U)|$, while the phase
\begin{equation}
\alpha(U) \equiv \frac{{\det}_{k}M(U)}{W_{|k|}(U)}
\label{eq:def_alpha}
\end{equation}
is introduced when computing the observable 
\begin{equation}
\left< O(U)\right>_{k} = \frac{\left<O(U) \alpha(U)\right>_{|k|}}{\left<\alpha(U)\right>_{|k|}},
\end{equation}
where $\left< \cdot\right>_{|k|}$ denotes the average with regard to
the generated ensemble based on the $W_{|k|}$ measure. In practice, to
evaluate the partition function numerically, the continuous Fourier
transform in eq.(\ref{eq:def_detk}) has so far either been replaced by
a discrete Fourier transform~\cite{Alexandru:2005ix} or by a more
sophisticated approximation~\cite{Meng:2008hj,Li:2010qf} and the so
introduced bias needs a careful treatment. With the reduced Wilson
fermion matrix these approximations have become obsolete, simply
because the Fourier transform can now be evaluated exactly.

Still, the quenching of the phase of the integration measure,
${\det}_{k} M(U) \rightarrow W_{|k|}$, introduces a systematic error
which one needs to control. A measure of how severe the fluctuations
of the phase are, is provided by the expectation value of $\alpha(U)$
in eq.(\ref{eq:def_alpha}) in a given ensemble. It turns out that the
sign fluctuations for the canonical ensembles seem to be under good
control and one might wonder how generic this feature is. One way to
estimate the reliability of the simulations is to reweigh results
generated at one value of $k$ to other values $k'$ and to check
consistency between the reweighted results and the ones obtained from
direct simulations. For the reweighting from one canonical ensemble
$Z_C(k)$ to another $Z_C(k')$ the relevant quantity is
\begin{equation}\label{eq:reweighting_factor}
\alpha_{|k|\rightarrow k'}(U)= \frac{{\det}_{k'}M(U)}{W_{|k|}(U)} \, . 
\end{equation}
For this definition of the reweighting factor we see from the second
column of Table~\ref{tab:phaseav1} that its magnitude changes very
fast as we move away from the original ensemble, in this case the one
with $n_B=4$.  However, the value of the factor changes for all
configurations in a similar manner and the average is still
comfortably away from zero in terms of its variance (cf.~third and
fourth column) even though its magnitude is dramatically
changed. (Note that this dramatic change is related to the variation
of $\det_k M$ with $k$ over many orders of magnitude.)  What we mean
by that is that its magnitude, as compared with its variance, is
larger than two or more. In principle it is when this ratio becomes
close to one that the reweighting in the equation above will run into
numerical difficulties. We see from Table~\ref{tab:phaseav1} that for
baryon numbers as large as $n_{B}'=16$ the average is still twice
the variance.
\begin{table}[!th]
   \centering
   \begin{tabular}{@{} lrrr @{}} 
      \toprule
    $n_{B}$ & $\left<\alpha\right> \quad\,\,\,\,$ & $\sigma_{\alpha} \quad\,\,\,\,$ & $\sigma_{\alpha}/\left<\alpha\right>$ \\
      \midrule
00 & $7.63\cdot 10^{+2}$ & $3.93\cdot 10^{+2}$ & $0.515\,\,\,$\\
$04^{*}$ & $4.87\cdot 10^{-1}$ & $3.09\cdot10^{-2}$ & $0.064\,\,\,$\\
08 & $1.60\cdot 10^{-4}$ & $1.21\cdot 10^{-5}$ & $0.076\,\,\,$\\
12 & $7.78\cdot 10^{-9}$ & $1.11\cdot 10^{-9}$ & $0.143\,\,\,$\\
16 & $1.82\cdot 10^{-14}$ & $9.81\cdot 10^{-15}$ & $0.540\,\,\,$\\
      \bottomrule
   \end{tabular}
   \caption{The real part of the average reweighting factor
     eq.(\ref{eq:reweighting_factor}). The imaginary part vanishes by
     symmetry. This factor is computed based on the ensemble generated
     with $n_{B}=4$ (marked with a star above).}
   \label{tab:phaseav1}
\end{table}

\begin{figure}[!th]
   \centering
   \includegraphics[width=13cm]{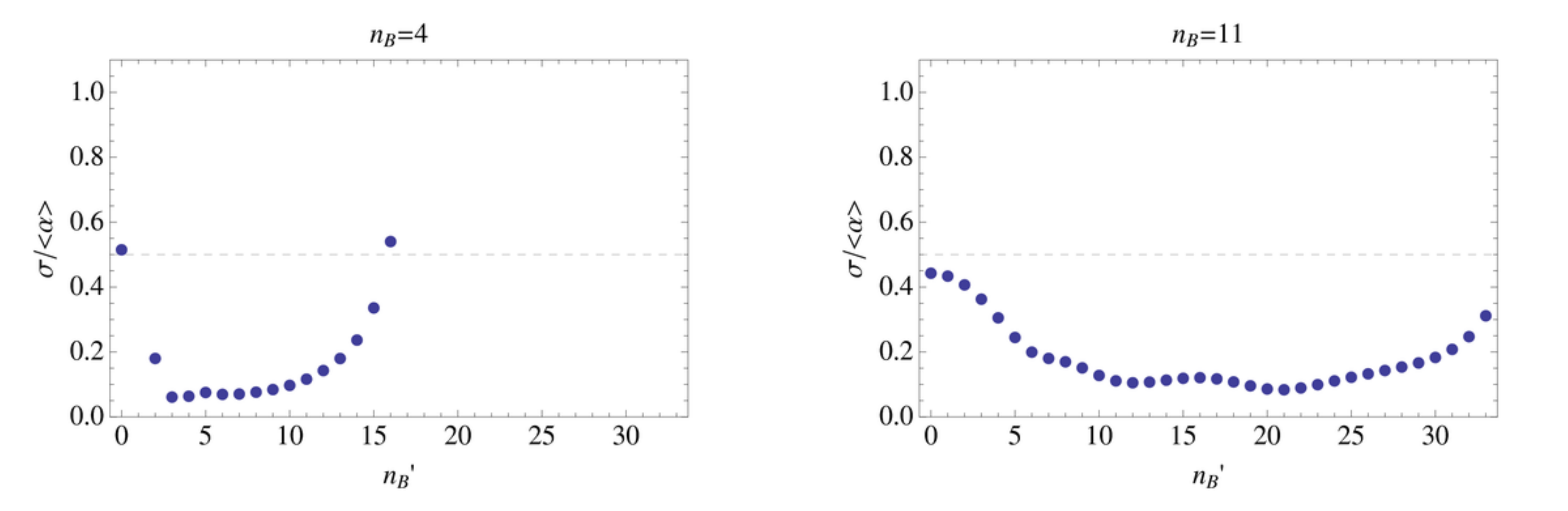}
   \caption{The real part of the average phase factor: the ratio between its
     variance and its mean value.}
   \label{fig:phaseav}
\end{figure}

In Fig.~\ref{fig:phaseav} we plot the ratio between the variance and
the average of the real phase factor reweighted from the canonical
ensembles with $n_B=4$ and 11, as a function of the reweighted baryon
number. Note that in both cases the average reweighting factor is well
behaved over a large range of reweighted baryon numbers.

\begin{figure}[!th]
   \centering
   \includegraphics[width=13cm]{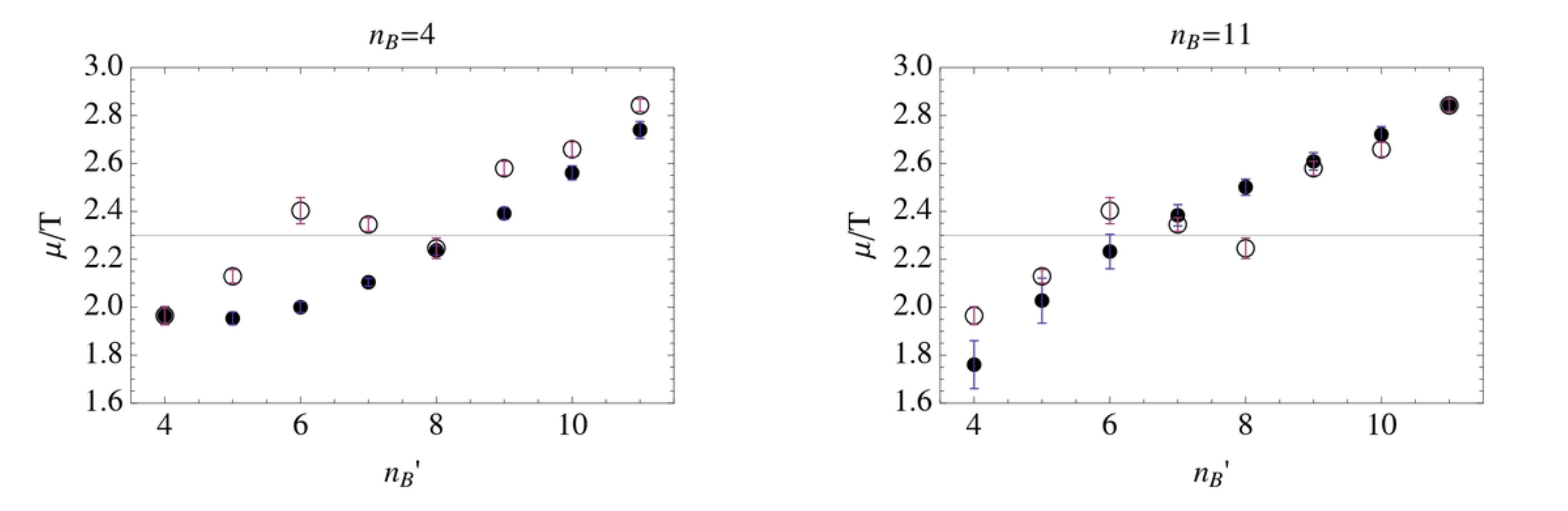}
   \caption{The baryon chemical potential reweighted from the ensemble
   with $n_B=4$ (left plot) and $n_B=11$ (right plot). Empty circles
   are the results of the direct calculation.}
   \label{fig:chempot}
\end{figure}

Encouraged by this result one might also try to reweight the chemical
potential as a function of the baryon number following the definition
in eq.(\ref{eq:muB_def}). The chemical potential, as defined there,
can be shown to be
\begin{eqnarray}
\mu_B(n_B)/T &=& -\log \frac{Z_{C}(3(n_B+1))}{Z_{C}(3n_B)} = 
-\log \left< \frac{{\det}_{3(n_B+1)} M}{{\det}_{3n_B} M}\right>_{3n_B} \\
&=&-\log \frac{\left<{\det}_{3(n_B+1)}M/W_{|3n_B|}\right>_{|3n_B|}}{\left< {\det}_{3n_B}M/W_{|3n_B|}\right>_{|3B|}}.
\label{eq:mu_canonical}
\end{eqnarray}
In Fig.~\ref{fig:chempot} we show the reweighted chemical potential as
defined by eq.(\ref{eq:muB_def}), for the same ensembles as before,
$n_B=4$ (left plot) and $n_B=11$ (right plot). Despite the absence of
a sign problem in the reweighting factor, it is evident that the
reweighting fails. The ensemble $n_B=4$ in the confined phase misses
to describe even the neigbouring ensemble with only slightly different
baryon number $n_B=5$. The ensemble $n_B=11$, on the other hand, is in
the deconfined phase and seems to be able to describe other deconfined
phases with different baryon numbers.  Not surprisingly, however, it
completely fails to describe ensembles with $n_B \leq 8$, and hence the
phase transition, simply because it does not contain any information
about the confined phase.

The next interesting question would now be to reweight an ensemble in
the mixed phase, e.g.~with $n_B=7$. However, the reweighted results
turn out to be too noisy, indicating that the information gathered in
the canonical ensembles is simply not enough to allow for a reliable
reweighting. One has to keep in mind that the canonical ensembles used
here only contain about 1500 configurations each. The next obvious
step is then to employ a multi-ensemble reweighting, combining the
information from all the ensembles into the reweighting. What we find
is that for $n_{B}'=4$ the most important contributions indeed come
from the ensembles with small $n_{B}$ and that, as we move towards
larger $n_{B}'$, higher ensembles come into play as expected. However,
even the multi-ensemble reweighting does not seem capable to reproduce
the S-shape behaviour typical for a first order phase transition, and
we conclude that the reweighting suffers from a severe overlap
problem.  Nevertheless, the example impressively demonstrates the ease
with which such calculations can be achieved using the reduced Wilson
fermion matrix.

\section{Conclusions}
We have presented a reduction method for Wilson Dirac fermions which
generates a dimensionally reduced fermion matrix. The size of the
reduced matrix is independent of the temporal lattice
extent. Moreover, the dependence of the matrix on the chemical
potential factors out and reduces to a simple multiplicative
factor. This allows to evaluate the Wilson fermion determinant for any
value of the chemical potential, once the eigenvalues of the reduced
matrix are calculated, and hence allows to perform the exact
projection of the determinant to the canonical sectors with fixed
fermion number.

The reduced fermion matrix presented here facilitates various
interesting applications, for example the reweighting of ensembles to
arbitrary values of the chemical potential or the fermion number. So
far, this has only been possible for staggered fermions. Another
application is the direct simulation of canonical ensembles and this is
now possible without any bias from inexact projections to the
canonical sectors.  Since the size of the reduced matrix is
independent of the temporal lattice extent, such calculations can in
principle be done at arbitrarily low temperatures, barring possible
sign problems.

The reduced fermion matrix has some interesting properties like the
spectral symmetry $\lambda \leftrightarrow 1/\lambda^*$, a simple
behavior of the spectrum under $\Z(N_c)$-transformations and the
correlation of the spectrum with the Polyakov loop. We believe that
such properties may be important for the development of more efficient
canonical simulation algorithms, or for the optimisation of reweighting
strategies.

As a first test we applied the reduction method to a set of canonical
ensembles and determined the phase fluctuations of the Wilson fermion
determinant at non-zero chemical potential, or non-zero fermion
number, using standard reweighting techniques. It turns out that for
the ensembles considered here, the overlap problem inherent in all
reweighting methods introduces a systematic bias that forbids reliable
calculations, e.g.~using multi-ensemble reweighting. On the other
hand, the phase fluctuations seem to be rather well controlled and it
will be interesting to see whether this is a generic feature of
canonical partition functions.

\subsection*{Acknowledgements}
We would like to thank Philippe de Forcrand for discussions, and
Atsushi Nakamura and Keitaro Nagata for correspondence on their
related work.  Andrei Alexandru is supported in part by
U.S. Department of Energy under grant DE-FG02-95ER-40907. The
computational resources for this project were provided by the George
Washington University IMPACT initiative.  We would like to thank Anyi
Li for providing us with the configurations needed for this study.

\bibliography{wilson_matrix_reduction}
\bibliographystyle{JHEP}
\end{document}